# Newtonian Potential in Quantum Regge Gravity

*Herbert W. Hamber*

Theory Division, CERN
CH-1211 Genève 23, Switzerland

*Ruth M. Williams*

Department of Applied Mathematics and Theoretical Physics
Silver Street
Cambridge CB3 9EW, England

**ABSTRACT**

We show how the Newtonian potential between two heavy masses can be computed in simplicial quantum gravity. On the lattice we compute correlations between Wilson lines associated with the heavy particles and which are closed by the lattice periodicity. We check that the continuum analog of this quantity reproduces the Newtonian potential in the weak field expansion. In the smooth anti-de Sitter-like phase, which is the only phase where a sensible lattice continuum limit can be constructed in this model, we attempt to determine the shape and mass dependence of the attractive potential close to the critical point in $G$. It is found that non-linear graviton interactions give rise to a potential which is Yukawa-like, with a mass parameter that decreases towards the critical point where the average curvature vanishes. In the vicinity of the critical point we give an estimate for the effective Newton constant.

# 1 Introduction

The lattice formulation presents a natural framework for determining the structure of nonperturbative effects in quantum gravity. Since Einstein gravity is not perturbatively renormalizable, the computation of radiative corrections in the weak field expansion around a flat metric cannot be controlled until at least a partial resummation of the perturbative series can be performed. Even then, contributions which are non-analytic in the coupling cannot be determined. From the analytical side there is some hope that an expansion in the coupling can be performed close to two dimensions, and thus provide some insight into the qualitative properties of the theory, while a numerical approach has the advantage that it can attack the four-dimensional case directly, without having to rely on an expansion in a small parameter.

Among the properties that should emerge from a consistent theory of quantum gravity one can list the recovery of almost flat space at large distances, and the appearance of an attractive Newtonian potential between heavy bodies. In a consistent lattice formulation of gravity the computation of the Newtonian potential is in principle no more difficult than the determination of the static potential in QCD. The Equivalence Principle could then in be tested by employing different sources for the gravitational field.

A crucial question, which has up to now only been partially addressed, is the existence of a lattice continuum limit. As in any lattice field theory, a continuum theory can only be recovered if the lowest lying excitation of the theory (the graviton) can be made to vanish, at least in some region of bare parameter space. It is only in this region that the details of the underlying lattice structure are washed out and the long distance universal properties of the continuum theory start to emerge. In this respect the correct excitation spectrum of the weak field expansion represents only a necessary, but not a sufficient requirement.

Even the existence of $a$ continuum limit by itself (whose appearance would be signaled by the presence of long wavelength fluctuations in coordinate invariant fluctuations and correlations) does not prove that General Relativity is recovered at large distance until one is able to show that the behavior of correlations is associated with a massless spin two particle. There is some hope though that if the action and measure have the correct symmetry properties, and if the correct states propagate in the weak field limit, then the same should be true in the full nonperturbative treatment of the theory. In this respect the determination of the Newtonian potential provides a crucial ingredient, since its long distance properties (combined with the Equivalence Principle) are characteristic of General Relativity.

Regge's formulation of gravity in terms of simplicial manifolds with varying edge lengths is the natural discretization for General Relativity [1]. At the classical level, it is the only lattice model known to reproduce in four dimensions General Relativity, with continuous curvatures, classical gravitational waves, and no graviton doubling problem in the weak field limit. The correspondence with continuum gravity is particularly transparent in the lattice weak field expansion, with the invariant edge lengths playing the role of infinitesimal geodesics in the continuum. In the limit of smooth manifolds with small curvatures, the continuous diffeomorphism invariance of the continuum theory is recovered [2, 3]. But in contrast to ordinary lattice gauge theories, the model is formulated entirely in terms of manifestly coordinate invariant quantities, the edge lengths, which form the elementary degrees of freedom in the theory [4, 2]. Of course in perturbation theory, the lattice theory remains non-renormalizable just as the continuum theory [5, 6]. This does not exclude the possibility that the theory might exist non-perturbatively, and well-known examples of such a behavior exist for simpler models both in the continuum and on the lattice [7].

Recent work based on Regge's simplicial formulation of gravity has shown in pure gravity the appearance in four dimensions of a phase transition in the bare Newton's constant, separating a *smooth* phase with small negative average curvature from a *rough* phase with large positive curvature. For sufficiently large higher derivative coupling the transition is continuous, with the curvature vanishing at the critical point with a universal exponent which has been determined to be approximately $\delta = 0.63(3)$ [8, 9]. While the fractal dimension seems rather small in the rough phase, indicating a tree-like geometry for the ground state, it is very close to four in the smooth phase close to the critical point. A calculation of the critical exponents in the smooth phase and close to the critical point seems to suggest that the transition is continuous (at least for sufficiently large higher derivative coupling) with divergent curvature fluctuations, and that a lattice continuum might therefore be constructed.

If the model has any resemblance to General Relativity at large distances, it should give rise to an attractive potential between heavy particles which should fall off like $1/r$, with subleading classical relativistic and quantum corrections. In general this is only expected to happen in the vicinity of the critical point at $G_c$, where the lattice continuum limit is to be taken, following the general prescription of Wilson for determining the low energy properties of quantum cutoff theories [10]. In the context of the weak-field expansion, the problem of determining the potential from the correlations of world-lines associated with two heavy particles has been discussed recently by Modanese in [11], and part of our work can be regarded as an extension to the non-perturbative case.

In this paper we will present some first qualitative result regarding the nature of the potential in simplicial gravity, as derived from numerical studies (on a lattice with $24 \times 16^4 = 1,572,864$ simplices), and will begin by considering the determination of the potential from the correlations of Wilson lines in the framework of the weak field expansion. The paper is organized as follows. In Sec. 2 we introduce the simplicial action and measure for the gravitational degrees of freedom. We then discuss the formulation and properties of Wilson line correlations and the potential in QED (Sec. 3) and quantum gravity, in the context of the continuum weak field expansion (Sec. 4) and on the lattice (Sec. 5). In Sec. 6 we present our results and in Sec. 7 some discussion. In Sec. 8 we discuss a simple mean field model for quantum gravity, and finally Sec. 9 contains our conclusions.

## 2   Action and Measure

We write the four-dimensional pure gravity action on the lattice as

$$I_g[l] = \sum_{hinges\ h} V_h \left[ \lambda - k\, A_h \delta_h / V_h + a\, A_h^2 \delta_h^2 / V_h^2 \right] , \qquad (2.1)$$

where $V_h$ is the volume per hinge (which is represented by a triangle in four dimensions), $A_h$ is the area of the hinge and $\delta_h$ the corresponding deficit angle, proportional to the curvature at $h$. All geometric quantities can be evaluated in terms of the lattice edge lengths $l_{ij}$, which uniquely specify the lattice geometry for a fixed incidence matrix (for a complete list of references on Regge gravity see for example [12]). The geometry is varied by varying the lengths of the edges, while the topology is fixed by assigning the incidence matrix [1]. The underlying lattice structure is chosen to be hypercubic, with a natural simplicial subdivision to ensure its overall rigidity [2, 13, 14, 15]. In the classical continuum limit the above action is then equivalent to

$$I_g[g] = \int d^4x \sqrt{g} \left[ \lambda - \tfrac{1}{2} k\, R + \tfrac{1}{4} a\, R_{\mu\nu\rho\sigma} R^{\mu\nu\rho\sigma} + \cdots \right] , \qquad (2.2)$$

with a bare cosmological constant term (proportional to $\lambda$), the Einstein-Hilbert term ($k = 1/8\pi G$), and a higher derivative term proportional to $a$ [16, 17, 18]. For an appropriate choice of bare couplings, the above lattice action is bounded below, due to the presence of the higher derivative term. In the continuum one finds that the action is bounded below for $a > 3k^2/8\lambda$, while for the regular tessellation of the four-sphere $\alpha_5$ represented by a 5-simplex one finds that the action is bounded below in the weak field expansion for $a > 0.471 \, k^2/\lambda$ [13].

---

[1] In the discrete dynamical triangulation model one keeps the edge lengths equal to one, and varies the incidence matrix. In this approach continuos diffeomorphism invariance is absent even for flat space. It is unclear if such models have a lattice continuum limit above two dimensions [30, 31].

In the quantum case, for non-singular measures and in the presence of the $\lambda$-term, a stable lattice can be shown to arise naturally for sufficiently small $k$ [14, 13, 15], thus allowing a non-perturbative definition of the Euclidean path integral. The higher derivative terms can be set to zero ($a = 0$), but they nevertheless may be necessary for reaching the lattice continuum limit [9], and are in any case generated by radiative corrections already in weak coupling perturbation theory. They are also present in the weak field expansion of the Regge-Einstein action.

The cosmological constant term with $\lambda > 0$ ensures that the volumes are bounded, while the measure prevents any of the edge lengths from becoming too small. Without loss of generality, one can set the bare cosmological constant $\lambda = 1$, in which case all lengths are measured in units of $\lambda^{-1/4}$. The theory then contains a natural ultraviolet cutoff, related to the average lattice spacing, $l_0 = \sqrt{<l^2>}$. It can be considered as a fundamental length scale [19], as an artificial device necessary in order to construct a lattice continuum limit, where it is sent to zero keeping physical quantities fixed, or as a quantity inherited from some more fundamental theory such as superstrings (where $l_0 = g\sqrt{\alpha'}$). We should add that since the model is formulated in a finite box, one does not expect any infrared divergences as long as the box size is finite. The box size can then be considered as an additional parameter which can be varied in order to study the renormalization properties of the theory [21].

The gravitational measure contains an integration over the elementary lattice degrees of freedom, the edge lengths. For the edges one writes the lattice integration measure as [13, 14, 15]

$$\int d\mu[l] = \prod_{edges\ ij} \int_0^\infty V_{ij}^{2\sigma} \, dl_{ij}^2 \, F[l] \ , \qquad (2.3)$$

where $V_{ij}$ is the 'volume per edge', $F[l]$ is a function of the edge lengths which enforces the higher-dimensional analogs of the triangle inequalities, and the power $\sigma = 0$ for the lattice analog of the DeWitt measure for pure gravity. The factor $V_{ij}^{2\sigma}$ plays a role analogous the factor $(\sqrt{g})^{2\sigma}$ which appears for continuum measures [22, 23]. A variety of measures have been proposed in the continuum [22, 23, 24, 25] and on the lattice [26, 27], some of which are even non-local. Since there is no exact gauge invariance on the Regge lattice away from smooth manifolds (nor in any other local lattice formulation of gravity), one cannot uniquely decide a priori which is the most appropriate gravitational measure. On the other hand the above measure integrates over the invariant degrees of freedom of the lattice theory, the edge lengths. Different gravitational measures which have been proposed differ only in the volume factors $\sqrt{g}$ appearing in the measure. We regard therefore the above measure as the most natural one on the Regge lattice.

We note that *no* cutoff is imposed explicitly on small or large edge lengths, if a non-singular measure such as $dl^2$ is used. We believe that this fact is essential for the recovery of diffeomorphism invariance close to the critical point, where on large lattices a few rather long edges, as well as some rather short ones, start to appear [9]. On the other hand an effective ultraviolet cutoff is generated dynamically, due to the presence of the cosmological constant term (at large $l$), and from the measure (at small $l$). This cutoff is of the order of the average edge length, $l_0 = \sqrt{<l^2>}$. We also note that no gauge fixing is necessary in this approach, since the volume of the diffeomorphism group, which appears for smooth enough manifolds, cancels out between numerator and denominator when invariant averages are computed. The influence of the measure and the dependence of the results on the underlying lattice structure have also been systematically investigated recently in [28], where a one-parameter family of measures has been introduced in the Regge formalism. The results seem to indicate that the effects of changing the measure are small for appropriately scaled physical quantities such as the average curvature, as long as the basic form of Eq. (2.3) is preserved, and in particular the generalized triangle inequality constraints.

## 3  Wilson loop and potential in QED

In an ordinary gauge theory such as QED and QCD the static potential can be computed from the Wilson loop [32]. To this end one considers the process where a particle-antiparticle pair (an electron and a positron in QED, a quark anti-quark pair in QCD) are created at time zero, separated by a fixed distance $R$, and re-annihilated at a later time $T$ (see Fig. 1.).

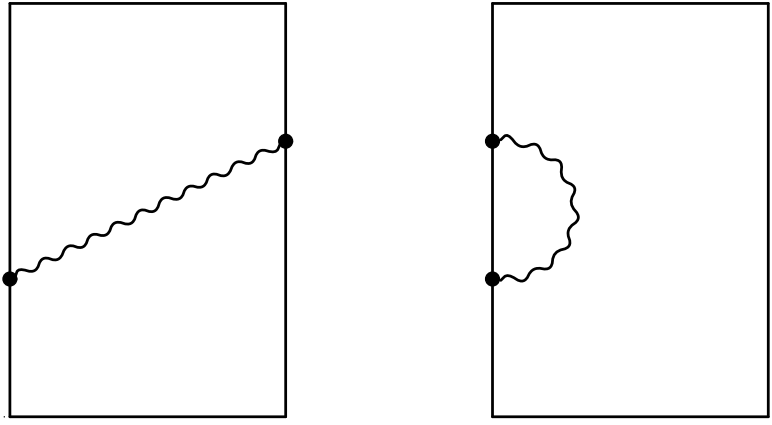

Fig 1. Square Wilson loop in QED.

In QED the amplitude for such a process associated with the closed loop $\Gamma$ is given by the Wilson loop

$$W(\Gamma) \; = \; < \exp\Big\{ ie \oint_\Gamma A_\mu(x) dx^\mu \Big\} > \; , \qquad (3.1)$$

which is a manifestly gauge invariant quantity. We recall here briefly the essential ingredients of the calculation in QED, in order to prepare for the perturbative quantum gravity computation in the next section. From the Euclidean QED action,

$$I(A) = \tfrac{1}{4} \int d^4x \; F_{\mu\nu}(x) F^{\mu\nu}(x) \; , \qquad (3.2)$$

one obtains the photon propagator in real space

$$\Delta_{\mu\nu}(x-y) \; = \; \frac{1}{4\pi^2} \frac{\delta_{\mu\nu}}{(x-y)^2} \; . \qquad (3.3)$$

If the calculation is done with a lattice cutoff, then the photon propagator at the origin is finite.[2] Since the integrals over the fields appearing in the QED Wilson loop are Gaussian, one gets immediately

$$< \exp\Big\{ ie \oint_\Gamma A_\mu dx^\mu \Big\} > \; = \; \exp\Big\{ -\tfrac{1}{2} e^2 \oint_\Gamma \oint_\Gamma dx^\mu dy^\nu < A_\mu(x) A_\nu(y) > \Big\} \qquad (3.5)$$

$$= \; \exp\Big\{ -\tfrac{1}{2} e^2 \oint_\Gamma \oint_\Gamma dx^\mu dy^\nu \Delta_{\mu\nu}(x-y) \Big\} \; . \qquad (3.6)$$

---

[2] On a hypercubic lattice one has

$$\int_{-\pi}^{\pi} \frac{d^4p}{(2\pi)^4} \frac{1}{4 \sum_\mu \sin^2 \frac{p_\mu}{2}} = 0.154933... \qquad (3.4)$$

Two types of contributions arise (from $x$ and $y$ on the same side versus opposite sides). These involve the two types of integral,

$$\int_\epsilon^T dy \int_0^T dx \frac{1}{(x-y)^2 + \epsilon^2} = \pi \frac{T}{\epsilon} - 2 \log \frac{T}{\epsilon} \;, \tag{3.7}$$

where $\epsilon \to 0$ is an ultraviolet cutoff of the order of the lattice spacing, and

$$\int_0^T dy \int_0^T dx \frac{1}{(x-y)^2 + R^2} = 2 \frac{T}{R} \arctan \frac{T}{R} - \log(1 + \frac{T^2}{R^2}) \;. \tag{3.8}$$

Adding all contributions together, and specializing to the case $T \gg R$, one gets

$$\oint_\Gamma \oint_\Gamma dx^\mu dy^\nu \Delta_{\mu\nu}(x-y) \} \simeq \frac{1}{2\pi\epsilon}(T+R) - \frac{1}{2\pi}\frac{T}{R} - \frac{1}{\pi^2}\log(\frac{T}{\epsilon}) \;, \tag{3.9}$$

and therefore for the Wilson loop itself

$$< \exp\{ie \oint_\Gamma A_\mu dx^\mu\} > \simeq \exp\{-\frac{e^2}{4\pi\epsilon}(T+R) + \frac{e^2}{4\pi}\frac{T}{R} + \frac{e^2}{2\pi^2}\log(\frac{T}{\epsilon}) + \cdots\} \tag{3.10}$$

$$\underset{T \gg R}{\sim} \exp\bigl[-V(R)\,T)\bigr] \;, \tag{3.11}$$

where use has been made of the fact that for large times the exponent in the amplitude involves the energy for the process times the time $T$. Then for $V(R)$ itself one obtains, up to a constant,

$$V(R) = -\lim_{T\to\infty} \frac{1}{T} \log < \exp\{ie \oint_\Gamma A_\mu dx^\mu\} > \sim \text{cst.} - \frac{e^2}{4\pi R} \;, \tag{3.12}$$

which is the correct Coulomb potential for two oppositely charged particles.

To obtain the potential it is not necessary to consider closed loops. Alternatively, in a periodic box one can introduce two long parallel lines in the time direction, separated by a distance $R$ and closed by the periodicity of the lattice, and associated with oppositely charged particles,

$$< \exp\{ie \int_\Gamma A_\mu dx^\mu\} \exp\{ie \int_{\Gamma'} A_\nu dy^\nu\} > \tag{3.13}$$

$$\simeq \exp\{-e^2 \int_\Gamma dx^\mu \int_{\Gamma'} dy^\nu < A_\mu(x) A_\nu(y) > -\tfrac{1}{2}e^2 \int_\Gamma \int_\Gamma \cdots - \tfrac{1}{2}e^2 \int_{\Gamma'} \int_{\Gamma'} \cdots\} > \;, \tag{3.14}$$

which gives

$$\underset{T \gg R}{\sim} \exp\{-\frac{e^2}{4\pi\epsilon}T + \frac{e^2}{4\pi}\frac{T}{R} + \frac{e^2}{2\pi^2}\log(\frac{T}{\epsilon}) + \cdots\} \sim e^{-TV(R)} \;, \tag{3.15}$$

and therefore the same result as before for the potential $V(R)$. This second setup is quite useful in practical applications in lattice QCD [33], and provides for an efficient and accurate method for computing the potential, since the time $T$ can be taken as large as the box size allows.

# 4 Gravitational Case - Perturbation Theory

In the gravitational case there is no notion of "oppositely charged particles", so one cannot use the closed Wilson loop to extract the potential [34]. One is therefore forced to consider a process in which two separate world-lines for the two particles are introduced.

It is well known that the free fall equation for a heavy spinless particle can be obtained by extremizing the space-time distance travelled [37]. The length of the geodesic connecting the two points is then

$$d_{min} = \min_{x_\mu(\tau)} d(a,b\,|\,g) \;, \tag{4.1}$$

where the distance along a path $x^\mu(\tau)$ between the points $a$ and $b$ in a fixed background geometry, characterized by the metric $g_{\mu\mu}$, is given by

$$d(a,b\,|\,g) = \int_{\tau(a)}^{\tau(b)} d\tau \sqrt{g_{\mu\nu}(x)\tfrac{dx^\mu}{d\tau}\tfrac{dx^\nu}{d\tau}} \;. \tag{4.2}$$

Thus the quantity

$$\mu \int_{\tau(a)}^{\tau(b)} d\tau \sqrt{g_{\mu\nu}(x)\tfrac{dx^\mu}{d\tau}\tfrac{dx^\nu}{d\tau}} \;, \tag{4.3}$$

where $\mu$ is the mass of the heavy particle, can be taken as the Euclidean action contribution associated with the heavy particle.

Next consider two particles of mass $\mu_1$, $\mu_2$, propagating along parallel lines in the 'time' direction and separated by a fixed distance $R$. We can consider space-time to be asymptotically flat in the time direction, but as we shall discuss below this is not necessary. We shall consider here a process of the type described in Fig. 2. Then the coordinates for the two particles can be chosen to be $x^\mu = (\tau, \pm R/2, 0, 0)$. The amplitude for this process is a product of two factors, one for each heavy particle [11]. Each is of the form

$$L(0;\,\mu_1) = \exp\Big\{-\mu_1 \int d\tau \sqrt{g_{\mu\nu}(x)\tfrac{dx^\mu}{d\tau}\tfrac{dx^\nu}{d\tau}}\,\Big\} \;. \tag{4.4}$$

For the two particles we write the amplitude as

$$\text{Amp.} \equiv W(0,R;\,\mu_1,\mu_2) = L(0;\,\mu_1)\,L(R;\,\mu_2) \;. \tag{4.5}$$

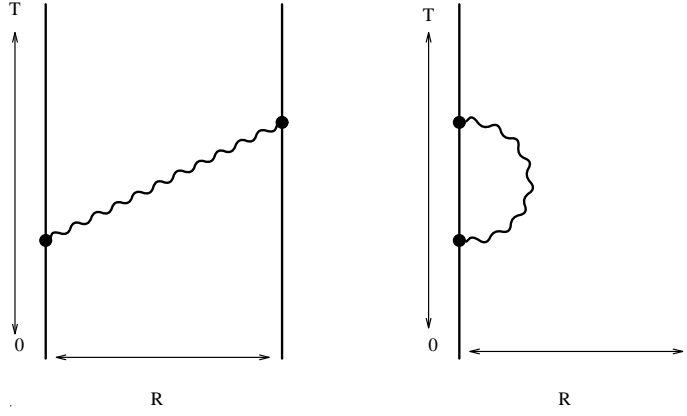

Fig 2. Worldlines for two heavy particles at rest and lowest order graviton exchanges.

For weak fields we set $g_{\mu\nu} = \delta_{\mu\nu} + h_{\mu\nu}$, with $h_{\mu\nu} \ll 1$, and therefore $g_{\mu\nu}(x)\frac{dx^\mu}{d\tau}\frac{dx^\nu}{d\tau} = 1 + h_{00}(x)$. Then for above geometry (two parallel worldlines) the amplitude reduces to

$$W(\mu_1,\mu_2) = \exp\left\{-\mu_1 \int_0^T d\tau \sqrt{1 + h_{00}(\tau)}\right\} \exp\left\{-\mu_2 \int_0^T d\tau' \sqrt{1 + h_{00}(\tau')}\right\} . \quad (4.6)$$

Expanding the square roots,

$$e^{-T\mu_1} \exp\left\{-\tfrac{1}{2}\mu_1 \int_0^T d\tau\, h_{00}(\tau) + \cdots\right\} e^{-T\mu_2} \exp\left\{-\tfrac{1}{2}\mu_2 \int_0^T d\tau'\, h_{00}(\tau') + \cdots\right\} , \quad (4.7)$$

and factoring out the metric-independent rest mass contribution one has

$$\sim e^{-(\mu_1+\mu_2)T} \left\{1 + \tfrac{1}{4}\mu_1\mu_2 \int_0^T d\tau \int_0^T d\tau'\, h_{00}(\tau)h_{00}(\tau') + \cdots\right\} . \quad (4.8)$$

After averaging over the $h_{\mu\nu}$ field (with $<h_{\mu\nu}> = 0$) one obtains

$$<W(\mu_1,\mu_2)> \; = e^{-(\mu_1+\mu_2)T} \left\{1 + \tfrac{1}{4}\mu_1\mu_2 \int_0^T d\tau \int_0^T d\tau' <h_{00}(\tau)h_{00}(\tau')> + \cdots\right\} . \quad (4.9)$$

In momentum space the graviton propagator, in the DeWitt-Feynman gauge $\partial_\mu h_{\mu\nu} = 0$, is given by [38]

$$\Delta_{\alpha\beta\mu\nu}(k) = 16\pi G\, \frac{\delta_{\alpha\mu}\delta_{\beta\nu} + \delta_{\alpha\nu}\delta_{\beta\mu} - \delta_{\alpha\beta}\delta_{\mu\nu}}{k^2} , \quad (4.10)$$

and therefore in real space

$$<h_{\alpha\beta}(x)h_{\mu\nu}(y)> \; = \frac{4G}{\pi}\, \frac{\delta_{\alpha\mu}\delta_{\beta\nu} + \delta_{\alpha\nu}\delta_{\beta\mu} - \delta_{\alpha\beta}\delta_{\mu\nu}}{(x-y)^2} . \quad (4.11)$$

In our case we just need

$$<h_{00}(\tau)h_{00}(\tau')> \; = \frac{4G}{\pi}\, \frac{1}{(\tau-\tau')^2 + R^2} , \quad (4.12)$$

and the averaged amplitude then becomes

$$e^{-(\mu_1+\mu_2)T} \left\{ 1 + \mu_1\mu_2 \frac{G}{\pi} \int_0^T d\tau \int_0^T d\tau' \frac{1}{(\tau-\tau')^2 + R^2} + \cdots \right\} , \qquad (4.13)$$

or, since $G$ is assumed to be small,

$$\sim \exp\left\{ -(\mu_1+\mu_2)T + \mu_1\mu_2 \frac{G}{\pi} \int_0^T d\tau \int_0^T d\tau' \frac{1}{(\tau-\tau')^2 + R^2} + \cdots \right\} . \qquad (4.14)$$

The integrals are easily evaluated,

$$\int_0^T d\tau \int_0^T d\tau' \frac{1}{(\tau-\tau')^2 + R^2} = 2\frac{T}{R} \arctan\frac{T}{R} - \log(1+\frac{T^2}{R^2}) \underset{T \gg R}{\sim} \pi \frac{T}{R} - 2\log\frac{T}{R} + \cdots , \qquad (4.15)$$

and thus the averaged amplitude is given by

$$< W(0,R; \mu_1,\mu_2) > = \exp\left\{ -T \left( \mu_1 + \mu_2 - G \frac{\mu_1\mu_2}{R} \right) + \cdots \right\} . \qquad (4.16)$$

Since the amplitude gives, for large times, the energy $E$ for the state, $<\text{Amp.}> \sim \exp(-ET)$, one finds that the potential has indeed the expected form, $V(R) = -G\,\mu_1\mu_2/R$. Incidentally we note that, had we done the calculation in $d$ dimensions, we would have obtained for the coefficient of the $R$-dependent part $2(d-3)/(d-2)R^{3-d}$ which vanishes, as expected, in $d=3$ [39].

The contribution involving the sum of the two particle masses is $R$ independent, and can be subtracted, if the Wilson line correlation is divided by the averages of the individual single line contribution. For *one* particle one has to lowest order in the weak field expansion

$$< L(0; \mu_1) > \equiv < \exp\left\{ -\mu_1 \int d\tau \sqrt{g_{\mu\nu}(x)\frac{dx^\mu}{d\tau}\frac{dx^\nu}{d\tau}} \right\} > \underset{T \gg R}{\sim} e^{-\mu_1 T} . \qquad (4.17)$$

One can then compute the correlation between (closed) Wilson lines of length $T$, separated by an average distance $R$, and extract the Newtonian potential from

$$V(R) = -\lim_{T \to \infty} \frac{1}{T} \log \frac{< W(0,R; \mu_1,\mu_2) >}{< L(0; \mu_1) >< L(R; \mu_2) >} \sim -G\frac{\mu_1\mu_2}{R} . \qquad (4.18)$$

If one is only interested in the spatial dependence of the potential, one can simplify things a bit and take the two masses to be equal, $\mu_1 = \mu_2 = \mu$.

To higher order in the weak field expansion one has to take into account multiple graviton exchanges [40], contributions from graviton loops and self-energy contributions to the heavy particles. The first two modify the shape of the Newtonian potential, while the latter has the effect of renormalizing the mass of the heavy particles which enter in the potential. According to the Equivalence Principle, one would then expect the potential to involve these effective, renormalized masses only [3]. To see this effect, it is instructive to compute the average of *one* Wilson line,

$$\exp\left\{-\mu \int_0^T d\tau \sqrt{1+h_{00}(\tau)}\right\} , \tag{4.19}$$

for which the lowest order diagrams are shown in Fig. 3.

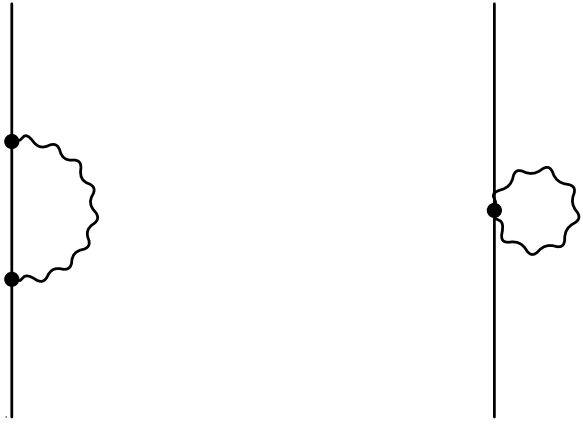

Fig 3. Lowest order graviton exchange contributions to the Wilson line.

Expanding again the square root,

$$e^{-\mu T} \exp\left\{-\tfrac{1}{2}\mu \int_0^T d\tau\, h_{00}(\tau) + \tfrac{1}{8}\mu \int_0^T d\tau\, h_{00}^2(\tau) + \cdots \right\} , \tag{4.20}$$

one gets

$$\sim e^{-\mu T} \left\{ 1 - \tfrac{1}{2}\mu \int_0^T d\tau\, h_{00}(\tau) + \tfrac{1}{8}\mu \int_0^T d\tau\, h_{00}^2(\tau) + \tfrac{1}{8}\mu^2 \int_0^T d\tau \int_0^T d\tau'\, h_{00}(\tau) h_{00}(\tau') + \cdots \right\}. \tag{4.21}$$

One then averages over the $h_{\mu\nu}$ field ($<h_{\mu\nu}> = 0$), using the graviton propagator given previously.

Next one needs the regulated integral ($\epsilon \to 0$)

$$\int_0^T d\tau \int_0^T d\tau' \frac{1}{(\tau-\tau')^2 + \epsilon^2} = 2\,\frac{T}{\epsilon} \arctan\frac{T}{\epsilon} - \log(1+\frac{T^2}{\epsilon^2}) \underset{T \gg \epsilon}{\sim} \frac{\pi}{\epsilon} T - 2\log\frac{T}{\epsilon} , \tag{4.22}$$

and the expectation value then becomes

$$e^{-\mu T} \left\{ 1 + \tfrac{1}{8}\,\mu T\,\frac{4G}{\pi\epsilon^2} + \tfrac{1}{8}\,\mu^2\,\frac{4G}{\pi}\left[\frac{\pi}{\epsilon}T - 2\log\frac{T}{\epsilon}\right] + \mathcal{O}(G^2) \right\} , \tag{4.23}$$

or

$$\begin{aligned} <L(0;\,m)> &= \exp\left\{-\mu T\,[1 - \frac{G}{2\pi\epsilon^2} - \frac{\mu G}{2\epsilon} + \mathcal{O}(\frac{\log T}{T})]\right\} \\ &\sim \left(\frac{\epsilon}{T}\right)^{\mu^2 G/\pi} e^{-\tilde\mu T} , \end{aligned} \tag{4.24}$$

---

[3] We thank P. Menotti for a discussion on this point.

where we have introduced the effective mass $\tilde{\mu}$,

$$\tilde{\mu} = \mu \left( 1 - \frac{G}{2\pi\epsilon^2} - \frac{\mu G}{2\epsilon} + \cdots \right) . \qquad (4.25)$$

A partial resummation of the perturbation expansion can be done without having to rely on the weak field expansion. Introduce the operator associated with the exponent of one Wilson line operator

$$\mathcal{L}_\Gamma = \int_\Gamma d\tau \sqrt{g_{\mu\nu}(x)\frac{dx^\mu}{d\tau}\frac{dx^\nu}{d\tau}} , \qquad (4.26)$$

where $\Gamma$ is the path associated with the heavy particle. We have paths in mind that are close or equal to geodesic and are very long (of lengths comparable to the box size) and separated from each other by a large distance. Then we can write

$$< e^{-\mu_1 \mathcal{L}_{\Gamma_1}} e^{-\mu_2 \mathcal{L}_{\Gamma_2}} > = \qquad (4.27)$$

$$< (1 - \mu_1 \mathcal{L}_{\Gamma_1} + \frac{1}{2!}\mu_1^2 \mathcal{L}_{\Gamma_1}\mathcal{L}_{\Gamma_1} + \cdots)(1 - \mu_2 \mathcal{L}_{\Gamma_2} + \frac{1}{2!}\mu_2^2 \mathcal{L}_{\Gamma_2}\mathcal{L}_{\Gamma_2} + \cdots) > , \qquad (4.28)$$

or

$$< 1 - \mu_1 \mathcal{L}_{\Gamma_1} - \mu_2 \mathcal{L}_{\Gamma_2} + \mu_1\mu_2 \mathcal{L}_{\Gamma_1}\mathcal{L}_{\Gamma_2} + \frac{1}{2!}\mu_1^2 \mathcal{L}_{\Gamma_1}\mathcal{L}_{\Gamma_1} + + \frac{1}{2!}\mu_2^2 \mathcal{L}_{\Gamma_2}\mathcal{L}_{\Gamma_2} + \mathcal{O}(\mu^3) > . \qquad (4.29)$$

Next we write the part that does not involve correlations between the lines $\Gamma_1$ and $\Gamma_2$ as

$$1 - \mu_1 < \mathcal{L}_{\Gamma_1} > + \frac{1}{2!}\mu_1^2 < \mathcal{L}_{\Gamma_1}\mathcal{L}_{\Gamma_1} > + \cdots \simeq e^{-\tilde{\mu}_1 T} , \qquad (4.30)$$

which should be valid if the path $\Gamma_1$ is very long. We shall also assume here that the two very long paths have comparable lengths $T$. Here $\tilde{\mu}_1 = \mu_1 + \delta\mu_1$ is the effective, renormalized mass. Then the whole expression above in Eq. (4.29) can be factored as

$$(1 - \mu_1 < \mathcal{L}_{\Gamma_1} > + \frac{1}{2!}\mu_1^2 < \mathcal{L}_{\Gamma_1}\mathcal{L}_{\Gamma_1} > + \cdots) \times$$
$$(1 - \mu_2 < \mathcal{L}_{\Gamma_2} > + \frac{1}{2!}\mu_2^2 < \mathcal{L}_{\Gamma_2}\mathcal{L}_{\Gamma_2} > + \cdots) \times$$
$$(1 + \mu_1\mu_2 < \mathcal{L}_{\Gamma_1}\mathcal{L}_{\Gamma_2} > - \mu_1\mu_2 < \mathcal{L}_{\Gamma_1} >< \mathcal{L}_{\Gamma_2} > + \cdots) , \qquad (4.31)$$

which one can exponentiate

$$\simeq \exp\{-\tilde{\mu}_1 T\} \exp\{-\tilde{\mu}_2 T\} \exp\{+ \tilde{\mu}_1\tilde{\mu}_2 < \mathcal{L}_{\Gamma_1}\mathcal{L}_{\Gamma_2} >_c + \cdots\} , \qquad (4.32)$$

where $< \cdots >_c$ denotes the connected correlation. Higher order terms will then involve triple correlations of the type $< \mathcal{L}_{\Gamma_1}\mathcal{L}_{\Gamma_1}\mathcal{L}_{\Gamma_2} >$. In the front of the last correlation we have also replaced $\mu$ by $\tilde{\mu}$. Thus

$$T\, V(r) = -\tilde{\mu}_1\tilde{\mu}_2 \{< \mathcal{L}_{\Gamma_1}\mathcal{L}_{\Gamma_2} > - < \mathcal{L}_{\Gamma_1} >< \mathcal{L}_{\Gamma_2} >\} + \cdots , \qquad (4.33)$$

where $r$ is some average separation between the two particle paths. This last equation shows that the potential itself is related to the connected line-line correlation function. If the correlation is positive, then the potential should be attractive. The above expansion shows therefore the correspondence between the potential and the connected correlation between line operators. In the weak field expansion it of course just reproduces the result obtained previously, namely

$$\begin{aligned} T\ V(r) &= -\mu_1\mu_2 \left\{ < \int_0^T d\tau\sqrt{1+h_{00}(\tau)} \int_0^T d\tau'\sqrt{1+h_{00}(\tau')} > \right. \\ &\quad \left. - < \int_0^T d\tau\sqrt{1+h_{00}(\tau)} > < \int_0^T d\tau'\sqrt{1+h_{00}(\tau')} > \right\} \\ &= -\mu_1\mu_2 \left\{ T^2 + \tfrac{1}{4} \int_0^T d\tau \int_0^T d\tau' < h_{00}(\tau)h_{00}(\tau') > + \cdots - T^2 - \cdots \right\} \\ &= -T\ \mu_1\mu_2\ \frac{G}{r} \end{aligned} \qquad (4.34)$$

## 5  Gravitational Case - Lattice Theory

At this point, the prescription for computing the Newtonian potential for quantum gravity should be clear. For each metric configuration (which is a configuration of edge lengths on the lattice) one chooses a geodesic that closes due to the lattice periodicity (and there might be many that have this property for the topology of a four-torus), with length $T$. One then enumerates all the geodesics that lie at a fixed distance $R$ from the original one, and computes the associated correlation between the Wilson lines. After averaging the Wilson line correlation over many metric configurations, one extracts the potential from the $R$ dependence of the correlation of Eq. (4.18). Indeed, by this method it should be even possible to check for homogeneity and isotropy of the underlying random lattice.

On the lattice one can construct the analog of the Wilson line for one heavy particle,

$$L(x,y,z) \;=\; \exp\{-\mu \sum_i l_i\} \;, \qquad (5.1)$$

where edges are summed in the "$t$" direction, and the path is *closed* by the periodicity of the lattice in the $t$ direction. Since we envision the simplicial lattice as divided up in hypercubes according to the prescription of Ref. [2], the points $x, y, z$ can be taken as the remaining labels for the Wilson line.

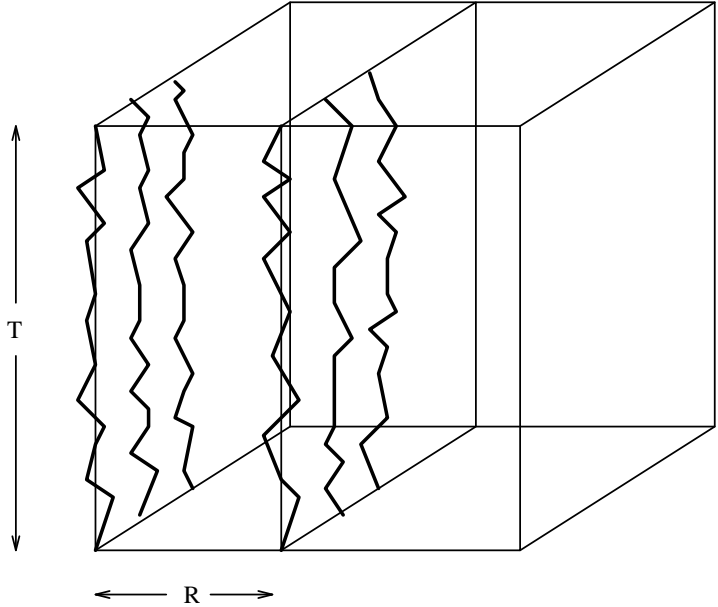

Fig 4. Correlations between Wilson lines closed by the lattice periodicity.

For a single line we expect

$$< L(x,y,z) > = < \exp\{-\mu \sum_i l_i\} > \sim e^{-\tilde{\mu}T} , \qquad (5.2)$$

where $T$ is the linear size of lattice in the chosen $t$ direction, $T = <V>^{1/4}$, where $<V>$ is the average volume of the space-time lattice. The correlation between Wilson lines at average "distance" $R$ is then given by

$$-\frac{1}{T} \log \left[ \frac{< L(x,y,0)\, L(x,y,R) >}{< L(x,y,0) >< L(x,y,R) >} \right] \underset{T \gg R}{\sim} V(R) . \qquad (5.3)$$

In practice it is better to assume that for large $R \gg l_0$ the potential has the form

$$V(R) \underset{R \gg m^{-1}}{\sim} - G(R)\, \mu_1 \mu_2 \, \frac{e^{-mR}}{R} , \qquad (5.4)$$

corresponding to a Yukawa potential, allowing for the possibility of a small graviton "mass" $m$. This is suggested by the fact that in anti-de Sitter space the graviton propagator has an exponential tail at large distances, which should reflect itself in the behavior of the potential [41, 42]. And in fact the "smooth phase" of lattice gravity, which is the only physically acceptable phase in this model, has $<R> < 0$ up to the critical point at $G_c$ [9]. Classically, the characteristic "mass" appearing in this case is related to the non-vanishing scalar curvature $R < 0$ of anti-de Sitter space,

$$m = 1/a_0, \quad \text{with} \quad R = -12/a_0^2 . \qquad (5.5)$$

This happens in spite of the fact that *no* explicit mass is given to the graviton, and therefore presumably no Ward identities need to be violated in the quantum case (a similar situation arises in three-dimensional gravity, where the transverse-traceless mode (the graviton) can acquire a mass without violating gauge invariance [43] ). On the other hand such a behavior should not be unexpected given the presence of the infrared cutoff $a_0$ that appears in an anti-de Sitter space.

A similar result is found in the weak field expansion around flat space [38, 45], where the presence of a cosmological term gives rise to a "mass" for $\lambda < 0$,

$$m^2 = -2\lambda/k = -R/2 \ , \tag{5.6}$$

although arguments based on the weak field expansion about flat space in the presence of a cosmological constant should be taken with care, due to the presence of the tadpole term, linear in the weak field $h_{\mu\nu}$. For de Sitter space ($R > 0$), it is known that no such mass term can arise, and in fact it has been argued recently that (Minkowski) de Sitter space is inherently unstable [44].

In the anti-de Sitter case the Einstein equations for the vacuum become

$$-\partial^2 g_{\mu\nu} - 2\Lambda g_{\mu\nu} = 0 \ , \tag{5.7}$$

with $\Lambda$ related to the Ricci scalar via $R = 4\Lambda = 4\lambda/k$. Thus for negative scalar curvature the mass is real. The range associated with the potential is then $\sim \hbar/(mc)$. In the real world this number must be very small. From the fact that super-clusters of galaxies apparently do form, one can set a limit on the range, $> 10^{25} cm$, or $m < 10^{-30} eV$ [45].

# 6  Numerical Results

Let us now discuss the numerical methods employed in this work and the analysis of the results. As in our previous work, the edge lengths are updated by a straightforward Monte Carlo algorithm, generating eventually an ensemble of configurations distributed according to the action of Eq. (2.1) and measure of Eq. (2.3). Further details of the method as applied to pure gravity are discussed in [14, 9], and will not be repeated here. In this work the edge length configurations already generated in [9] were used as a starting point.

For computing the potential, we considered lattices of size $16 \times 16 \times 16 \times 16$ (with 65536 sites, 983040 edges, 1572864 simplices). Even though these lattices are not very large, one should keep in mind that due to the simplicial nature of the lattice there are many edges per hypercube with many interaction terms, and as a consequence the statistical fluctuations can be comparatively small, unless measurements are taken very close to a critical point, and at rather large separation in the case of the potential. The results we present here are rather preliminary, and in the future it should be possible to repeat such calculations with improved accuracy on a much larger lattice.

As usual the topology is restricted to a four-torus (periodic boundary conditions). We have argued before that one could perform similar calculations with lattices employing different boundary conditions or topology, but the universal infrared scaling properties of the theory should be determined only by short-distance renormalization effects. The renormalization group equations are in fact expected to be independent of the boundary conditions, which enter only in their solution as it affects the correlation functions through the presence of a new dimensionful parameter, the linear system size $L = <V>^{1/4}$.

In this work the bare cosmological constant $\lambda$ appearing in the gravitational action of Eq. (2.1) was fixed at 1 (this coupling sets the overall scale in the problem), and the higher derivative coupling $a$ was set to 0 (pure Regge-Einstein action). For the measure of Eq. (2.3) this choice of parameters leads to a well behaved ground state for $k < k_c \approx 0.060$ for $a = 0$ [9, 28]. The system then resides in the 'smooth' phase, with a fractal dimension close to four; on the other hand for $k > k_c$ the curvature becomes very large ('rough' phase), and the lattice tends to collapse into degenerate configurations with very long, elongated simplices [14, 13, 15]. For $a = 0$ we investigated six values of $k$ $(0.00, 0.01, 0.02, 0.03, 0.04, 0.05)$. The case $a = 0$, which we have chosen to analyze first, represents the simplest situation, where explicit higher derivative terms are absent. In the future we plan to investigate the behavior of the potential for $a$ small but nonzero, and in particular in the regime $a > 3k^2/8\lambda$, where the Euclidean action is bounded below in the continuum.

From physical considerations it seems reasonable to impose the constraint that the scale of the curvature in magnitude should be much smaller than the average lattice spacing, but much larger than the size of the system, or in other words

$$<l^2> \quad \ll \quad <l^2> |\mathcal{R}|^{-1} \quad \ll \quad <V>^{1/2} \ . \tag{6.1}$$

This corresponds to the statement that in momentum space the physical scales should be much smaller that the ultraviolet cutoff, but much larger than the infrared cutoff. It also corresponds to the fact that in ordinary lattice field theory we usually require

$$L^{-1} \quad \lesssim \quad m \quad \lesssim \quad l_0^{-1} \ , \tag{6.2}$$

where $L$ is the linear size of the system, $m$ a typical mass, and $l_0$ the lattice spacing.

This fact prevents us from studying values of $k$ close to the critical point $k_c$, where the curvature becomes small and the correlation length (or inverse graviton mass) becomes larger than the system size. Conversely, far away from $k_c$ the curvature becomes rather large in magnitude, and the results become sensitive to the details of the ultraviolet cutoff. The above constraint then requires that $k$ be rather close, but not too close, to $k_c$, so as to be located within the "scaling window" of Eq. (6.2), where results relevant for the continuum theory should hopefully be obtained.

Another source of error comes from the fact that on a finite lattice there will be fluctuations in the critical value of $k$, $k_c$. We have considered lattices where the number of degrees of freedom is of order $10^6$. The energy density is not fixed, and there are fluctuations of order $N^{-1/2}$. For $k$ close to $k_c$ in a rough approximation $k_c - k$ is proportional to the energy $E$, and one expects fluctuations in $k_c$ from configuration to configuration, with a Gaussian distribution and a width proportional to $\Delta E/E \sim N^{-1/2}$,

$$\mathcal{P}(k) \sim \exp\left[-A(k_c - k)^2 N/k_c^2\right] , \qquad (6.3)$$

where $A$ is some numerical coefficient. One must therefore stay in a region where

$$N_{conf} \ll \exp\left[A(k_c - k)^2 N/k_c^2\right] , \qquad (6.4)$$

where $N_{conf}$ is the number of configurations one is considering. This means in particular that one cannot get too close to $k_c$ on a small lattice, or otherwise one will encounter an instability [46].

On the $16^4$ lattice we generated 1100 consecutive configurations at $a = 0$, for each value of $k$. The results for different values of $k$ can be considered as completely statistically uncorrelated, since they originated from unrelated configurations. Results for a larger statistical sample are in progress and will be presented elsewhere.

We computed the potential following the method described in the previous sections, using several values for $k$ close to $k_c$. Before one computes the potential, a choice has to be made for the mass of the heavy particle $\mu$. In principle one would like to make $\mu$ as large as possible. On the other hand when $\mu$ is very large, the average of a single Wilson line becomes very small and one runs into numeric precision problems; for example for $\mu = 1$ the Wilson line on a $16^4$ lattice is of order $10^{-16}$. So one is forced to consider smaller values of $\mu$ such that they can be handled by the precision of the machine. We have tried initially three values for $\mu$, 0.5 0.25 and 0.125, and have found roughly consistent results for the scaled potential $V(r)/\mu^2$ (see discussion below). In the following (except in one case) we will use $\mu = 0.5$ for which we believe that double precision (<16 decimals) is adequate. For this choice of bare heavy mass the renormalization effects for the mass itself are rather small.

We find for all values which we have studied $\delta\mu \simeq -0.026$ to $-0.036$, with the renormalization effect increasing slightly towards the critical point. In the following we shall neglect such a small effect and present the results for the potential in scaled form by dividing by $\mu^2 = 0.25$.

Figs. 5 to 7 present our results for the potential. As discussed previously, the expectation is that the potential in the quantum theory close to the critical point should be attractive ($V(R) < 0$), that it should decrease like $1/r$ close to the ultraviolet fixed point at $G_c$, and that it should scale like $\tilde{\mu}^2$, for $\mu_1 = \mu_2 = \mu$. The first encouraging result is that close to the critical point the potential is indeed clearly attractive, $V(r) < 0$. At very short distances, comparable to one or two average lattice spacings, we expect the potential to show some oscillations due to the underlying lattice structure, and this is indeed what is observed, like in the case of the curvature-curvature correlation [47]. The oscillations could be reduced by using a larger bin width for the distance and averaging the potential within the bins, but then only few points would be left to display. This could be useful on a larger lattice. In fact, we have chosen to average the potential at distances of zero and one lattice spacing and present one single point at $r = l_0/2 \approx 1.18$, since at such short distance we expect to see mostly lattice artifacts. As usual the errors in the potential are estimated by using a standard binning procedure. For distances greater than 5 average lattice spacings ($r > 12$) the errors become quite large and we would need higher statistics to get useful results. Not unexpectedly, the potential is more difficult to determine at large distances, where it becomes small and tends to be drowned in the statistical noise. Also for $k < 0.03$ the potential becomes very small (which makes it difficult to measure accurately) and for $k$ close to zero it turns positive at large distances (corresponding to a repulsive potential). This is not completely unexpected, since, at least in the weak field expansion, the potential changes sign when $k < 0$. But of course the weak field expansion loses much of its validity when we move away from almost flat space, which corresponds to $k \simeq k_c$. Here we seem to find that this happens at a slightly larger value of $k \approx 0.02$. We will return to this issue later in the paper.

In Fig. 8 we show the heavy mass dependence for the potential as obtained at one value of $k$ and for a small statistical sample (100 configurations of the edge lengths), but using always exactly the same set of configurations for $\mu = 0.5, 0.25, 0.125$. As can be seen from the graph, the results are quite consistent with a $\mu^2$ dependence of the potential (if we fit the mass dependence to a power by averaging over all points at distance 0-14, we find that this power is about $1.94 \pm 0.40$, quite close to the expected value of 2).

To further analyze the behavior of the potential, one can attempt to fit it at 'large' distances, here meaning $r \gg l_0$, to an exponential decay, as indicated by the Yukawa form of Eq. (5.4),

$$V(r) = -c \frac{e^{-mr}}{r} . \qquad (6.5)$$

Alternatively, one can try to fit them to a power law close to the critical point at $k_c$

$$V(r) = -c \frac{1}{r^\sigma} . \qquad (6.6)$$

*If* the potential is fitted to an exponential decay, one finds that the behavior is consistent, close to $k_c$, with a small mass that decreases as one approaches the critical point. This is shown in Fig. 5. We clearly do not have at this point a lot of points which would allow us to give a precise estimate for this mass or its error. Close to this critical point let us write for the mass of the particle, which is expected to determine the long distance behavior of the potential,

$$m^2 \underset{k \to k_c}{\sim} A_p (k_c - k) . \qquad (6.7)$$

We find some evidence for a decrease in the mass towards the critical point, and for the amplitude we estimate $A_p = 1.09(60)$. Here we are making the implicit assumption that the mass will indeed go to zero at the same critical point. The results for the potential are certainly consistent with this assumption, but the accuracy of the results and the systematic errors associated with the fact that the distances $r$ are still rather small do not allow one yet to determine in a clean way if this is indeed what is happening. We will leave a more accurate determination of the mass parameters for future work.

The motivation for using the mass squared in the preceding equation is as follows. In our previous work we estimated the critical exponent $\nu$, which determines how the dynamical graviton mass approaches zero at the critical point, $m \sim (k_c - k)^\nu$, and found that it was close to $1/2$ (our best estimate, from Ref. [9] gave $\nu \simeq 0.41$ for $a = 0.005$). (Also it should be added for the sake of clarity, that the values we quote refer to 'physical' masses, and not to masses in units of the lattice spacing, which would be larger by about a factor of two, since, as we mentioned previously, the average lattice is not one, but about $l_0 \sim 2.36$).

Alternatively, we can plot the mass $m$ versus the average curvature. In general this procedure is quite useful since it avoids the problem of having to rely on an accurate determination of the critical point in $k$. Naively one would expect on the basis of dimensional arguments that

$$m^2 \underset{\mathcal{R} \to 0}{\sim} A_{pR} |\mathcal{R}| , \qquad (6.8)$$

but we cannot exclude that a non-trivial exponent appears in this case as well. Clearly again our results are not accurate enough at this point to determine the exponent with any accuracy. We shall return to the issue of the exponents later. Under the above assumption we estimate in this case $A_{pR} \simeq (0.06)^2$, which seems a rather small number. On the other hand one gets a number closer to one if one uses a more natural scale, the effective average anti-de Sitter radius (see Eq. 5.5), defined here by $a_0 = l_0\sqrt{12/|\mathcal{R}|}$), as a scale instead of the average curvature. We find here $m \simeq 0.49/a_0$. We should add that a hard breaking of diffeomorphism invariance should induce a graviton mass of the order of the ultraviolet cutoff, $m \sim \pi/l_0$, which at this point is inconsistent with all our results. On the other hand a first order transition cannot be excluded, where the ground state would become unstable before the mass (or the average curvature) reaches the value zero.

When the mass of the particle is rather small, it becomes difficult to distinguish an exponential decay from a pure power behavior. Close to the critical point one can fit the potential to a pure power instead, and one finds the quality of the fits to be comparably good (for a comparison see for example Fig. 6). In Fig. 11 this effective power is plotted versus $k$, and one finds that it is somewhat greater than one, reflecting the fact that the potential falls off more rapidly in distance as one moves away from the critical point. From Fig. 11 we estimate the power at the critical point to be about $\sigma = 0.99(68)$ (the smallest power we actually measure at $k = 0.05$ is about 1.67, so we get the smaller values only by following the general trend and extrapolating to $k \simeq 0.60$).

If we exclude from the potential the point at $r = 1.2$ which corresponds more or less to the "origin", one finds that the decrease in distance $r$ is not very far from a $1/r$ behavior. In Fig. 7 we show a fit to the potential which is purely $1/r$ for $r > l_0$, and it seems that also this fit is rather good close to the critical point. This would give further support to the claim that the potential is very close to $1/r$ in the vicinity of the critical point, with some small mass or other correction. A radical possibility would be that the mass is actually zero, but this would seem unlikely in the presence of an average negative curvature, and would be at variance with the fact that the curvature-curvature correlation appears to be exponentially decaying close to the critical point [47]. At the present moment our results are not sufficiently accurate to determine inequivocably what those corrections are, and we can only give estimates for the size of the corrections given an assumed form. Needless to say, if we try to fit the potential to a function with more than two parameter such as $-c\exp(-mr)/r^\sigma$, we run into the problem of not having enough statistically significant points to constrain the parameters sufficiently.

In conclusion, our first results are not inconsistent with the expectation that close to the critical point the potential between heavy particles should be proportional to the mass squared of the particles, and that it should decreases like $1/r$ at short distances. A careful study of the above issues should give further support to the argument that coordinate invariance is indeed recovered in this model at large distances, and that the correct low energy theory is recovered in the vicinity of the fixed point.

# 7 Discussion

It is of interest to extract the effective Newton constant in the vicinity of the critical point. In general we expect that the Newton constant will depend on the distance $r$, and so we should write $G(r)$ for it. Furthermore, we should take into account the fact that all our dimensionful quantities are measured in units of some unit cutoff (it was set to one in Eq. (2.1)), and that our average lattice spacing $l_0$ is not quite equal to one (this is a small effect). At short distances $r \sim l_0$ we measure the coupling at scales close to the ultraviolet cutoff, while at larger distance we should see some renormalization effects, if they are there (Some time ago in a very nice paper the short distance behavior of pure Einstein gravity was discussed, exploiting the invariance of the classical Einstein action under dilatations [48]). Since we only have a few points in $r$ for the potential at any given $k$, we will restrict here our attention to the behavior of $G$ at short distances, close to the fixed point. Let us define here $G_{eff} = c$ as the coefficient of the potential obtained from the three fitting procedures used previously ($-c \exp(-mr)/r$, $-c/r^\sigma$, and $-c/r$). In the end we shall only be interested in the values in the close vicinity of the critical point.

As a function of $k$, the three sets of coefficients are shown in Fig. 12. One notices, not unexpectedly, that the values for $G_{eff}$ defined in the above way start to differ significantly as one moves away from the critical point, a reflection for example of the fact that the assumption of almost pure $1/r$ behavior is only valid in the vicinity of the critical point, and possibly only at rather short distances. On the other hand all three estimates seem to converge more or less to one value at $k_c$, which we estimate to be about 0.14 if we look at the results in Fig. 12. It is certainly encouraging that the value for the effective Newton constant at short distances in the vicinity of the fixed point is not zero or infinite in lattice units. Both values are close to the bare value, $G_c = 0.63$. Indeed the effective Newton constant we computed contains necessarily the cutoff, so we can write $G_{eff} = 0.15 = G_0 (l_0/s)^2$, where $G_0$ is comparable to $G_c$ and $s$ is a number of order one (it is $\pi$ if we use a momentum cutoff on a regular hypercubic lattice). In our case a discrepancy between $G_c$ and $G_0$ can be resolved by taking $s \simeq 4.84$.

We should keep in mind that even at the critical point where the curvature vanishes the lattice is by no means regular, and $l_0 = \sqrt{<l^2>}$ only represents an "average" cutoff. We should also perhaps recall here the fact that a bare cosmological constant $\lambda$, which could appear in the original action (as indicated in Eq. (2.1)) has been scaled out, when we set it equal to one by rescaling all the edge lengths. If we put it back in, then the effective Newton's constant would have to be multiplied by that scale, $G_{eff} = G_0 (l_0/s)^2 / \sqrt{\lambda}$, and $G_0$ and $s$ are the numbers discussed previously. As far as the distance dependence of the coupling $G(r)$ is concerned, we have nothing to say, based on our results so far on the potential. Of course if the potential decreases exponentially at large distances, one should factor out this dependence before determining the distance dependence of the coefficient $G(r)$.

Let us now return to a discussion of the fact that the potential seems to vanish when $k$ gets close to $k = 0$. From our results in fact we estimate that as we move away from the fixed point the potential becomes very small close to $k = 0.02$, and turns repulsive beyond that value. If we look at the weak field expansion for the graviton propagator (see Eq. (4.11)), we see that there are two contributions of opposite sign, the one with the wrong (repulsive) sign being associated with the trace part $-\delta_{\alpha\beta}\delta_{\mu\nu}/x^2$ of the metric. In the Landau gauge a similar situation arises, since the graviton propagator contains two terms of opposite sign, one associated with the spin-2 part, and one associated with the spin-0 part, $\Delta(x) = [P^{(2)} - \frac{1}{2} P^{(0)}]/x^2$, where $P^{(2)}$ and $P^{(0)}$ are spin-2 and spin-0 projection operators [49]. Let us assume here that this description based on the weak field expansion is more or less reliable in the vicinity of the fixed point, where the average curvature is very small and (almost) flat space is recovered on the average.

But we know that as we approach the value $k = 0$ the Einstein term switches off and there cannot be any propagating gravitons (or their non-perturbative counterparts), at least for $a = 0$. The only remaining term in the action is the cosmological term, which contains no derivatives. On the other hand the lattice gravitational measure (Eq. (2.3)) contains a residual interaction between the volumes which is due to the generalized triangle inequality constraints. These constraints will be present for almost any sensible local lattice measure, irrespective of the detailed form of the overall volume factors that enter in it. The triangle inequality constraints will induce a residual interaction between the volumes and edges, which will be non-vanishing when $k = a = 0$. Indeed when the correlation function between volumes at fixed geodesic distance is computed directly, one finds that such a correlation is nonzero at $k = 0$ [47, 29], and the corresponding mass is about 0.44(3). Based on the previous discussion one would therefore *expect* that the potential should become repulsive in this case, since the spin-2 kinetic term in the action is completely absent in this limit.

A possible interpretation of our results for the potential is therefore the following: At $k = 0$ only the trace part of the metric propagates, and the potential is repulsive. Away from, but close to, $k = 0$ the spin-2 part starts to propagate, with a mass that is roughly $m \sim |\log k|/l_0$, since the amplitude for moving $n$ steps on the lattice is proportional to $k^n = \exp(-n|\log k|)$ in this limit. As we approach the fixed point $k \to k_c$ the spin-2 part starts to propagate over larger distances, since its mass is decreasing. The potential eventually turns attractive, as it should, and for $k$ close to $k_c$ the correct admixture of spin-2 and spin-0 is recovered as determined by general covariance for fluctuations in the vicinity of almost flat space. We should stress that there is no reason to expect that the spectrum of excitations will come out correctly at infinitely strong coupling ($k = 0$); after all this certainly does not happen even in lattice QCD. One would expect that the potential, as well as any other coordinate invariant correlation function, would start to scale properly only when the mass of the two particles (spin 0 and spin 2) becomes comparable, and in turn comparable to the natural curvature scale, $1/a_0 = \sqrt{|\mathcal{R}|/12}\, l_0^{-1}$. From the results on the potential, the correlations and the average curvature we estimate that at $k = 0.03$ these three scales become comparable $m \sim 1/a_0 \simeq 0.3$.

Let us now return to the issue of the critical exponents for gravity. In statistical field theory one associates the singularities in the thermodynamic functions and in the correlations with the divergence of a correlation length (or inverse mass) at the critical point [10]. In the lattice gravity case we can follow a similar line of reasoning. The natural candidate for the correlation length in the gravitational case is the inverse of the graviton mass, $m = \xi^{-1}$. Let us assume that the singular part of the free energy $\mathcal{F} = -V^{-1} \log Z$ scales like $\xi^{-d_H}$ where $d_H$ is a (perhaps fractal) dimension, which we expect to be close or identical to four. The first derivative with respect to $k$ of the log of the partition function should then scale like

$$\mathcal{R} \underset{k \to k_c}{\sim} -A_\mathcal{R} (k_c - k)^\delta , \qquad (7.1)$$

up to a constant (which we find to be zero, at least for sufficiently large $a$), with an exponent $\delta = d_H \nu - 1$ (Josephson scaling law), if we define the exponent $\nu$ by the usual relation [10]

$$m \underset{k \to k_c}{\sim} A_m (k_c - k)^\nu . \qquad (7.2)$$

The fluctuations in the curvature, obtained from the second derivative of the log of the partition function should in turn scale like

$$\chi_\mathcal{R} \underset{k \to k_c}{\sim} A_{\chi_\mathcal{R}} (k_c - k)^{\delta-1} . \tag{7.3}$$

The relationship expected on the basis of scaling, $\nu = (1 + \delta)/d_H$, also implies for a continuous phase transition where the curvature vanishes,

$$\mathcal{R} \underset{k \to k_c}{\sim} m^{d_H - 1/\nu} \sim m^{d_H \frac{d_H \nu - 1}{d_H \nu}} . \tag{7.4}$$

In Ref. [8, 9] the exponent $\delta$ was estimated, in the presence of a small higher derivative term ($a = 0.005$ in Eq. (2.1)) to control the fluctuations in the curvature, at about $\delta = 0.63(3)$, which then gives $\nu d_H = 1.63$, and for the power in Eq. (7.4) about $0.39 \times d_H$. For $a = 0$ a smaller value was found, but with a much larger error, $\delta = 0.0 - 0.3$. A variety of methods can be used in principle to determine accurately the values of the critical exponents (such as direct determinations, finite size scaling [10, 21, 9], and real-space renormalization group methods based on block-spin ideas [13]).

Now if $d_H = 4$ then we get $\nu = 0.41(1)$, in which case the power appearing in Eq. (7.4) would be 1.55 [4]. In principle $\nu$ and therefore $d_H$ could be determined either directly from Eq. (7.2) or from Eq. (7.4), but our results so far are not sufficiently accurate to determine this power independently. It is amusing to note that if $\mathcal{R} \sim m^2$ (as assumed in Eq. (6.8), see also Fig. 10.) then $d_H - 1/\nu = 2$, which would imply a fractal dimension slightly above four, $d_H \sim 5.20$ and $\nu = 0.31$. (We also note that in this case the inverse mass $m$ becomes precisely (up to a constant) the anti-de Sitter radius, $m \sim a_0^{-1}$). To a certain extent we can exclude very large values for $d_H$, since these would imply (given the known value of $d_H \nu = 1.63(3)$) that the power in Eq. (7.4), $0.39 \times d_H$ is very large. But this does not seem the case if we look at Fig. 10. More accurate results would help in resolving this issue.

Let us recall here that a relationship like the one written in Eq. (7.1) and Eq. (7.2) is also suggested by the perturbative expansion for pure gravity about two dimensions. In the $2 + \epsilon$ perturbative expansion for gravity [51, 52] one analytically continues in the spacetime dimension by using dimensional regularization, and applies perturbation theory about $d = 2$, where Newton's constant is dimensionless. For the non-linear sigma model this is a completely sensible procedure, which gives reasonably accurate quantitative predictions in three dimensions [53]. It is not clear yet whether this approach makes sense for gravity beyond perturbation theory due to the unboundedness of the conformal mode, but it provides for a nice framework in which one can do controllable analytic calculations. In this expansion the dimensionful bare coupling is written as $G_0 = \Lambda^{2-d} G$, where $\Lambda$ is an ultraviolet cutoff (corresponding on the lattice to a momentum cutoff of the order of the inverse average lattice spacing, $\Lambda \sim 1/l_0$). A double expansion in $G$ and $\epsilon = d - 2$ then leads in lowest order to a nontrivial fixed point in $G$ above two dimensions. Close to two dimensions the gravitational beta function is given to one loop order by

$$\beta(G) \equiv \frac{\partial G}{\partial \log \Lambda} = (d-2)\,G - \beta_0\,G^2 + \cdots \;, \tag{7.5}$$

with $\beta_0 > 0$ for pure gravity. To lowest order the ultraviolet fixed point is then at $G_c = 1/\beta_0(d-2)$. Integrating Eq. (7.5) close to the non-trivial fixed point one obtains for $G > G_c$

$$m \;=\; \Lambda \, \exp\left(-\int^G \frac{dG'}{\beta(G')}\right) \underset{G \to G_c}{\sim} \Lambda\,|G - G_c|^{-1/\beta'(G_c)} \sim \Lambda\,|G - G_c|^{1/(d-2)} \;, \tag{7.6}$$

where $m$ is an arbitrary integration constant, with the dimensions of a mass, and which should be associated with some physical scale. It would appear natural here to identify it with the graviton mass, or the scale of the average curvature. The derivative of the beta function at the fixed point defines the critical exponent $\nu$, which to this order is independent of $\beta_0$,

$$\beta'(G_c) \;=\; -(d-2) \;=\; -1/\nu \;. \tag{7.7}$$

---

[4] It is amusing to note that a similar value ($\nu = 0.401$) was found, using real space renormalization group methods, in the Abelian $U(1)$ lattice gauge theory in four dimensions [50]. We thank G. Parisi for reminding us of this result.

The possibility of algebraic singularities in the neighborhood of the fixed point, appearing in vacuum expectation values such as the average curvature and its derivatives (Eq. (7.1) and Eq. (7.2), is then a natural one, at least from the point of view of the $2 + \epsilon$ expansion.

The previous results also illustrate how in principle the lattice continuum limit should be taken [10]. It corresponds to $\Lambda \to \infty$, $G \to G_c$ with $m$ held constant; for fixed lattice cutoff the continuum limit is approached by tuning $G$ to $G_c$. Alternatively, one can choose to compute dimensionless ratios directly, and determine their limiting value as one approaches the critical point. Away from $G_c$ one will in general expect to encounter some lattice artifacts, which reflect the non-uniqueness of the lattice transcription of the continuum action and measure, as well as its reduced symmetry properties.

In four dimensions we define the exponent $\nu$ by

$$m \underset{G \to G_c}{\sim} \Lambda\,|G - G_c|^\nu \;, \tag{7.8}$$

where $m$ is proportional to the graviton mass. Knowing $\nu$ is then equivalent to knowing $\beta'(G_c) = -1/\nu$. The value of $\nu$ determines the running of the effective coupling $G(\mu)$, where $\mu$ is an arbitrary momentum scale. The renormalization group tells us that in general the effective coupling will grow or decrease with length scale $1/\mu$, depending on whether $G > G_c$ or $G < G_c$, respectively. For $G > G_c$, corresponding to the smooth phase, one expects

$$G(\mu) = G_c + \left(\frac{m}{\mu}\right)^{1/\nu} + \mathcal{O}\left(\left(\frac{m}{\mu}\right)^{2/\nu}\right) . \tag{7.9}$$

There are indications from the lattice theory that only the smooth phase with $G > G_c$ exists (in the sense that spacetime collapses onto itself for $G < G_c$), which would suggest that the gravitational coupling can only increase with distance, as indicated by Eq. (7.9) [9].

Let us digress on possible corrections to the above formulae, which we have in general no reason to exclude. Let us assume that close to the ultraviolet fixed point at $G_c$ one can write the following expansion

$$\beta(G) = -\tfrac{1}{\nu}(G - G_c) - c\,(G - G_c)^2 + \mathcal{O}((G - G_c)^3) , \tag{7.10}$$

We are assuming here that at least the beta function is analytic at $G_c$, which is usually the case. After integrating as before, one finds for the structure of the correction

$$\left(\frac{m}{\Lambda}\right)^{1/\nu} = (G - G_c) - c\,\nu\,(G - G_c)^2 + \mathcal{O}((G - G_c)^3) . \tag{7.11}$$

The hope of course is that these corrections are small ($c \ll 1$), at least in the vicinity of the fixed point; the higher order term is unimportant if $(G - G_c) \ll 1/(c\nu)$. For the effective running coupling one then has the corresponding relation

$$G(\mu) = G_c + \left(\frac{m}{\mu}\right)^{1/\nu} + c\,\nu\,\left(\frac{m}{\mu}\right)^{2/\nu} + \mathcal{O}\left(\left(\frac{m}{\mu}\right)^{3/\nu}\right) . \tag{7.12}$$

One cannot exclude in principle more pathological behavior. If the leading term in the beta function in the vicinity of the fixed point vanishes,

$$\beta(G) = -c\,(G - G_c)^\sigma + \cdots , \tag{7.13}$$

(with $\sigma > 1$), one obtains an essential singularity in the mass gap,

$$\frac{m}{\Lambda} = \exp\{-\frac{1}{c(\sigma - 1)}(G - G_c)^{1-\sigma}\} . \tag{7.14}$$

It is not clear what should be the mechanism for such a cancellation in gravity, but if we consider such a possibility then one obtains instead of a power law a logarithmic scaling for the effective coupling (similar to what happens in QCD),

$$G(\mu) = G_c + \left[ c(\sigma - 1) \log \frac{\mu}{m} \right]^{-1/(\sigma-1)} . \tag{7.15}$$

But we should point out that this does not seem to happen in the $2 + \epsilon$ expansion, nor is there any evidence that it happens in the lattice model, but for now one should leave such a possibility open. We would like to add that even in the flat case one does not have in the Regge case anything resembling a regular lattice, although, contrary to lattices with random coordination number [54], the coordination number here stays fixed. It is known that already for flat random lattices novel critical behavior can arise, under certain conditions [55, 56]

The mass $m$ determines the size of scaling corrections, and plays therefore a role similar to $\Lambda_{\overline{MS}}$ in QCD. It cannot be determined perturbatively (as it appears here as an integration constant). It separates the short distance, ultraviolet regime with characteristic momentum scale $\mu \gg m$, or, more precisely, since we have an ultraviolet cutoff,

$$l_0^{-1} \gg \mu \gg m , \tag{7.16}$$

from the large distance, infrared region

$$m \gg \mu \gg L^{-1} , \tag{7.17}$$

where $L = <V>^{1/4}$ is the linear size of the system.

In quantum gravity it is of great interest to try to determine the value of the low energy, renormalized coupling constants, and in particular the effective cosmological constant $\lambda(\mu)$ and the effective Newton's constant $G(\mu) = 1/(8\pi k(\mu))$. Equivalently, one would like to be able to determine the large distance limiting value of a dimensionless ratio such as $\lambda(\mu)G^2(\mu)$, and perhaps even its dependence on the linear size of the system $L = V^{1/4}$ (which is another parameter in the model). (In the real world one knows that at laboratory scales $G_{eff} = (1.6160 \times 10^{-33} cm)^2$, while $\lambda_{eff} G_{eff}^2 \sim 10^{-120}$ is very small). In the continuum, these issues have been addressed in the context of Feynman diagram perturbation theory [57].

If $\nu$ is positive, then the beta function has a negative slope at the fixed point. This seems to be the case in the lattice theory. The increase or decrease in coupling as a function of scale is determined by what phase one is in. But on the lattice only the smooth phase is found to have an apparently sensible continuum limit. One immediate consequence of this result is that in the smooth phase with $G > G_c$ the gravitational coupling constant $G$ must increase with distance (anti-screening), at least for rather short distances. The opposite behavior (screening) would be true in the phase with $G < G_c$, but such a phase is known not to be stable and leads to no lattice continuum limit [9]. On purely physical grounds one would expect gravity to anti-screen (since it couples to everything with the same sign), and it is therefore not surprising that in the lattice theory the rough phase, where the opposite would be true, is pathological.

In conclusion, one then obtains for the dimensionful Newton's constant the following scale dependence, valid for short distances, $\mu \gg m$,

$$G(\mu) \underset{\mu \gg m}{\sim} l_0^2 \, \lambda^{-1/2} \left[ G_c + \left( \frac{m}{\mu} \right)^{1/\nu} \right] \;, \qquad (7.18)$$

(where $G_c$ is a pure number and $1/\nu \simeq 2.46$ if $d_H = 4$). Here again $l_0$ is of the order of the average lattice spacing, and we have restored the correct dimensions for $G(\mu)$ (length squared) and re-introduced the bare cosmological constant $\lambda$, which was previously set to one in Eq. (2.1) (it only sets the overall length scale).

As discussed in [8], the vacuum expectation value of the scalar curvature can be used as a definition of the effective, long distance cosmological constant,

$$\mathcal{R} \sim \frac{<\int \sqrt{g}\,R>}{<\int \sqrt{g}>} \sim \left( \frac{4\lambda}{k} \right)_{eff} \;. \qquad (7.19)$$

One can also introduce a classical anti-de Sitter radius $a_0$, by setting $|\mathcal{R}| = 12 l_0^2/a_0^2$. If the curvature vanishes at $k_c$ (see Eq. (7.1)) this radius diverges at $k_c$, and thus $(\lambda/k)_{eff} \to 0$ in lattice units. The exponent $\delta$, which is expected to be universal, was estimated previously to be about $\delta \simeq 0.63$ [8, 9]. The standard scaling arguments discussed previously then tell us that $\delta$ and $\nu$ are related via $\delta = d_H \nu - 1$, where $\nu$ is the correlation length exponent appearing in Eq. (7.8), and $d_H$ is the effective dimension of space (here close to four).

A more suitable definition of the running cosmological constant $\lambda(\mu)$ is as follows. Introduce a sphere $\Omega$ of size $r$, and compute the magnitude of the average curvature within that region,

$$\mathcal{R}_{\Omega(r)} \sim \frac{<|\int_{\Omega(r)} \sqrt{g}\,R|>}{<\int_{\Omega(r)} \sqrt{g}>} \;. \qquad (7.20)$$

At short distances (small spheres) the curvature fluctuates wildly and $\mathcal{R}_{\Omega(r)}$ is of the order of the ultraviolet cutoff, $\sim l_0^{-2}$. At larger distances (larger spheres) the curvature decreases, since the fluctuations tend to average out, and $\mathcal{R}_{\Omega(r)}$ approaches some average curvature value $\mathcal{R}_0$, which is determined by the chosen values for the bare parameters $k$, $\lambda$ and $a$ chosen in Eq. (2.1). Thus away from the critical point one expects

$$\mathcal{R}_{\Omega(r)} \sim \mathcal{R}_0 + c\, l_0^{-2}\, e^{-mr} \; , \tag{7.21}$$

whereas very close to the critical point, where both $\mathcal{R}_0$ and $m$ should go to zero, we expect that the exponential decay should turn into a power law decay

$$\mathcal{R}_{\Omega(r)} \sim l_0^{-2} \, (l_0/r)^\gamma \; , \tag{7.22}$$

with an exponent $\gamma = \delta/\nu = d_H - 1/\nu$. Thus for the cosmological constant itself we obtain

$$\lambda(\mu) \underset{\mu \gg m}{\sim} l_0^{-4} \, (l_0 \mu)^{d_H - 1/\nu} \lambda \, [1 + \mathcal{O}(m/\mu)] \; , \tag{7.23}$$

(with again $d_H - 1/\nu \simeq 1.54$ if $d_H = 4$), and we have restored the correct dimensions for $\lambda(\mu)$ (inverse length to the fourth power). For the dimensionless ratio $\lambda G^2$ one then obtains, from Eqs. (7.18) and (7.23),

$$(\lambda G^2)(\mu) \underset{\mu \gg m}{\sim} G_c^2 \, (l_0 \mu)^{d_H - 1/\nu} \, [1 + \mathcal{O}(m/\mu)] \; . \tag{7.24}$$

In conclusion, it seems that the dimensionless product $G^2 \lambda$ can be made very small, provided the momentum scale $\mu$ is small enough, or, in other words, at sufficiently large distances. We should add also that the fixed point value for the dimensionless gravitational constant, $G_c$, is in general non-universal, and depends on the specific way in which an ultraviolet cutoff is introduced in the theory (here via an average lattice spacing). In our model it is of order one for very small $a$, but for larger $a$ it decreases in magnitude. It would be of course of some interest to determine the scale dependence of the average curvature $\mathcal{R}_{\Omega(r)}$, and verify directly the behavior described above. Alternatively, one could study the behavior of deficit angles associated with large loops. Since the average curvature becomes very small close to the critical point. One would expect these deficit angles (which correspond to physical processes in which coordinate vectors are parallel transported around large, macroscopic loops) to be rather small.

How can one fix the fundamental lattice spacing $l_0^2$ (or $l_0^2/\sqrt{\lambda}$, if $\lambda$ is not equal to one in the original action) in this model? While there are apparently large fluctuations in the curvature at short distances, these fluctuations tend, as we said, to average out at large distances, if one is sufficiently close to the fixed point (otherwise the interactions are short ranged, and there is no noticeable gravitational potential). At such large distances it seems reasonable to assume that the only surviving contribution to the macroscopic energy $E$ is represented by the average curvature. Within a very large region of size $a_0$ the macroscopic action is then given only by the $R$ term, $E a_0 = -(16\pi G)^{-1} \int \sqrt{g} R$. Let us estimate this contribution. The integral should be restricted to a region of size $a_0$, since the gravitational interaction apparently falls off exponentially beyond distances of the order of $a_0$. Thus $E a_0 \approx G^{-1}(a_0) a_0^4 \times a_0^{-2}$, and since $G(a_0) \sim l_0^2$ one obtains $E \approx l_0^{-2} a_0$. In other words, the macroscopic energy only grows linearly with size. Solving for $l_0$, one obtains an estimate for the lattice cutoff $l_0^2 \approx a_0/E$, and for Newton's constant at "short" distances, $r \ll a_0$, $G \approx G_c a_0 / E$, where $G_c$ is a dimensionless number of order one.

Let us add that the larger $G_c$, the smaller the distance dependence of $G(r)$, since one has for the distance variation the (lowest order) result

$$\frac{\delta G(r)}{G(r)} = \frac{\nu^{-1}}{G_c \, (mr)^{-1/\nu} + 1} \frac{\delta r}{r} \;, \qquad (7.25)$$

(we have set $r = 1/\mu$), so in practice $G_c$ cannot be too small, and $m$ has to be very small.

We conclude that a possible interpretation of our results up to now is that in this model the effective gravitational coupling close to the ultraviolet fixed point grows with distance. For the gravitational coupling our results suggest an infrared growth away from the fixed point of the type $G(\mu) \sim \mu^{-1/\nu}$, while for the cosmological constant we have found a decrease in the infrared, $\Lambda(\mu) \sim \mu^{d_H - 1/\nu}$, with an exponent $\nu$ given approximately by $\nu \simeq 0.41$ if $d_H \simeq 4$, and perhaps only weakly dependent on the matter content [58]. The scale that seems to separate the short from the long distance behavior is $m$, which should be very small close to the fixed point, of the order of the inverse anti-de Sitter radius $a_0$.

# 8 Mean Field Theory

In this section we will describe a simple mean-field approach to quantum gravity, which contains some (but not all) of the essential features observed in the numerical simulations. Write for the effective action (or effective potential) for the average curvature $\mathcal{R}$, neglecting the metric degrees of freedom entirely,

$$I_{eff}(\mathcal{R}) = (k_c - k) V \mathcal{R} + a V (-\mathcal{R})^\lambda \;. \qquad (8.1)$$

Classically one has of course $k_c = 0$, but fluctuations will give rise to a nonzero value for the critical coupling that separates the smooth ($k < k_c$) from the rough phase ($k > k_c$). The last term can be thought of parametrizing the lattice and continuum higher derivative terms, and the effects of radiative corrections, which also include the measure contribution. In the smooth phase of gravity $\mathcal{R} < 0$, so we can write $\mathcal{R} = -|\mathcal{R}|$ in this phase. As we mentioned, a physically acceptable phase with $\mathcal{R} > 0$ (rough phase) does not seem to exist [9]. Then

$$\frac{\partial I_{eff}}{\partial \mathcal{R}} = (k_c - k)V - a\lambda V \left(-\mathcal{R}\right)^{\lambda-1} , \qquad (8.2)$$

with stationary point at

$$\mathcal{R}_0 = -(a\lambda)^{-1/(\lambda-1)} \left(k_c - k\right)^{1/(\lambda-1)} , \qquad (8.3)$$

and we therefore identify the curvature critical exponent $\delta$ with $\delta = 1/(\lambda - 1)$. This in a sense justifies the original form for $I_{eff}$, since it is known that the average curvature is non-analytic at $k_c$, (see Eq. (7.1)), with $\delta$ universal, and $k_c$ and $A_\mathcal{R}$ dependent on $a$. The fluctuation in the curvature is then given by

$$\chi_{\mathcal{R}_0} = (a\lambda)^{1/(\lambda-1)} (\lambda - 1)^{-1} \left(k_c - k\right)^{-(\lambda-2)/(\lambda-1)} , \qquad (8.4)$$

with an exponent $\alpha = 2 - d_H \nu = (\lambda - 2)/(\lambda - 1)$, where $\nu$ is the correlation length exponent,

$$m \underset{k \to k_c}{\sim} A_m \left(k_c - k\right)^\nu , \qquad (8.5)$$

with $m$ the graviton mass and $d_H$ the effective dimension of space-time, which, as we mentioned, should be close or perhaps identical to the physical space-time dimension. Classically one has $\lambda = 1$ and therefore $\delta = 1$, but it is known that in 3 and 4 dimensions $\delta < 1$ [9]. As long as $\mathcal{R} < 0$ the above solution is stable, since

$$\frac{\partial^2 I_{eff}}{\partial \mathcal{R}^2} = +aV \lambda (\lambda - 1) (-\mathcal{R})^{\lambda-2} , \qquad (8.6)$$

$$= a V \lambda (\lambda - 1) (a\lambda)^{-(\lambda-2)/(\lambda-1)} \left(k_c - k\right)^{(\lambda-2)/(\lambda-1)} , \qquad (8.7)$$

which also requires $\lambda > 2$ ($\delta < 1$ or $\nu < d_H/2$) for the second derivative of $I_{eff}$ to be finite at the origin $\mathcal{R} = 0$. In this approach there is always only one minimum for $k < k_c$ and the transition can never be first order (which requires two non-degenerate minima). For $\mathcal{R} > 0$ the effective action is complex, as it should, since no stable ground state is found in the lattice theory for $\mathcal{R} > 0$. Two further predictions arise out of this model. The first one is that the amplitude of the average curvature should diverge when $a$ is small,

$$A_\mathcal{R} \sim a^{-1/(\lambda-1)} . \qquad (8.8)$$

(From the numerical results in four dimensions it is unclear whether this happens precisely for $a = 0$, close to the critical point in $k$). The second one is that the minimum becomes increasingly shallow as $a \to 0$, which can lead to large fluctuations in the average curvature, unrelated to the approach at the critical point at $k_c$. This is also apparently observed, since it has been quite difficult to extract the critical exponent $\delta$ when $a$ is very small (or zero). Indeed it is possible that the model becomes unstable close to the critical point when $a = 0$, and that the transition is first order in this case [9]. Of course one does not expect this mean field theory to be quantitatively accurate, just as it is not for scalar field theories in low dimensions. It only represents an effective theory for the curvature, which is represented here as a single scalar quantity, neglecting the metric degrees of freedom entirely.

# 9 Conclusions

In the previous sections we have presented some first results regarding the properties of the Newtonian potential in the context of a model for quantum gravity based on Regge's lattice formulation. We have proposed a method for determining the potential which is based on the computation of Wilson line correlations. We have shown that the Wilson line correlations give the expected result to lowest order in the weak field expansion. Later we have then presented some first numerical results which seem to indicate that the correct qualitative features of the potential should emerge close to the critical point. In particular it was found that the potential is attractive close to the critical point, in agreement with previous results which also indicated the presence of an attractive interaction between dynamical scalar particles [58]. Our numerical results have been rather limited since we investigated for simplicity only the case $a = 0$ (no explicit higher derivative terms), and we have not performed yet a systematic study of the lattice continuum limit for the potential. As for any correlation in gravity, the accurate determination of the potential as a function of distance is a difficult task, since at large distance the correlations are small and the statistical noise becomes large. Still, our preliminary results suggest that the potential has more or less the expected classical form in the vicinity of the critical point, both as far as the mass dependence and the distance dependence are concerned.

Away from the critical point our results suggest that the potential is Yukawa-like, with a "mass" that decreases with the average curvature. We have not been able to determine with any precision how this mass scales with the curvature as the curvature approaches zero. We have argued that the appearance of such a mass is natural in the quantum analog of Euclidean anti-de Sitter space, and is likely to be a consequence of the non-linear interactions of gravitons with a non-flat fluctuating background, and the presence of a natural infrared cutoff in an anti-de Sitter space. In any case a systematic study of the potential should provide one more quantitative handle on the approach to the lattice continuum limit: the mass associated with the potential has to scale to zero close to the critical point in order for the theory to describe gravity. Based on previous work, where curvature fluctuations were found to diverge close to the continuous critical point, there is hope that this will happen when the accuracy of the present calculations will be improved.

We have not been able to determine in this work the distance dependence of the effective Newton's constant, although we expect on the basis of the phase diagram and the values of the critical exponents that in the smooth phase with $G > G_c$ gravitational interactions will increase slowly with distance. We have argued that the scale for such deviations from scale independence is set by the average curvature, which is very small close to the fixed point. Let us add that it would be very interesting to compute the Newtonian potential in three dimensions, where the leading spatial dependence is expected to be logarithmic, but with a vanishing coefficient (for zero cosmological constant).

**Acknowledgements**

We wish to thank Stanley Deser, Pietro Menotti, Giampiero Paffuti, Giorgio Parisi and Gabriele Veneziano for discussions and useful comments on this work. H.W.H thanks the Theory Division at CERN for hospitality and support during the completion of this work. The numerical computations were performed at NCSA under a *Grand Challenge* allocation grant. The authors wish to thank the director Larry Smarr for continuous support of this project. The parallel version of the program used in this work was written for the CM5 in collaboration with Y. Tosa of TMC, and his invaluable help is here gratefully acknowledged. This work was supported in part by the US National Science Foundation under grant PHY-9208386 and by the UK Science and Engineering Research Counsel under grant GR/J64788.

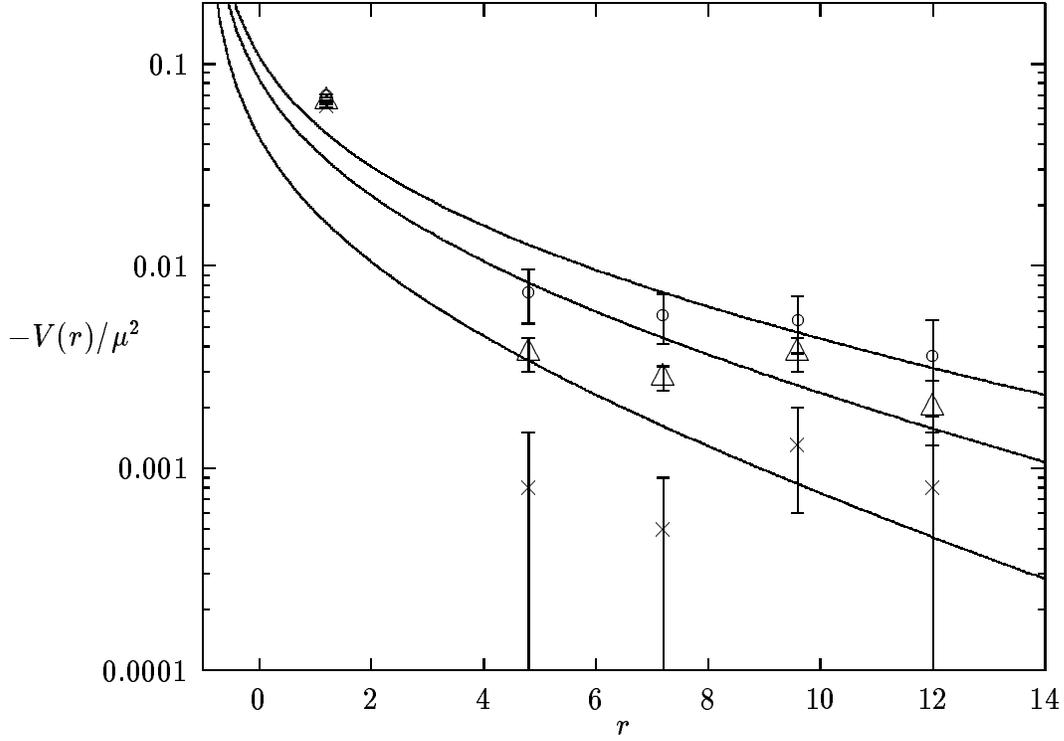

Fig. 5 . Computed scaled potential $-V(r)/\mu^2$ for $\lambda = 1$ and $a = 0$, and $k$=0.03 ($\times$), 0.04 ($\triangle$), and 0.05 ($\circ$). ($k_c \approx 0.060$). The lattice has $16^4$ sites, and the average lattice spacing for this range of parameters is $l_0 \simeq 2.36$. The lines represent best fits to the data of the form $c\exp(-mr)/r$.

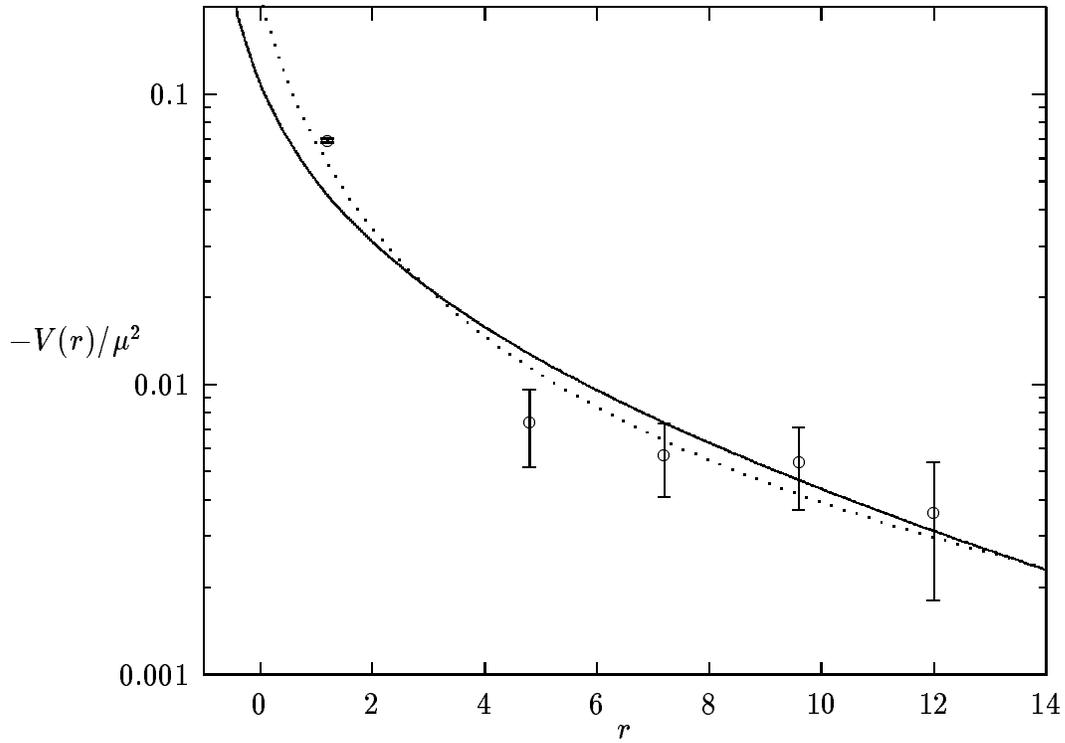

Fig. 6 . Potential $V(r)$ for $k = 0.05$ ($\circ$) only. The continuous lines represent best fits to the data of the form $c \exp(-mr)/r$ (with $m = 0.12$), while the dotted lines represent fits to $c/r^\sigma$ (with $\sigma = 1.67$).

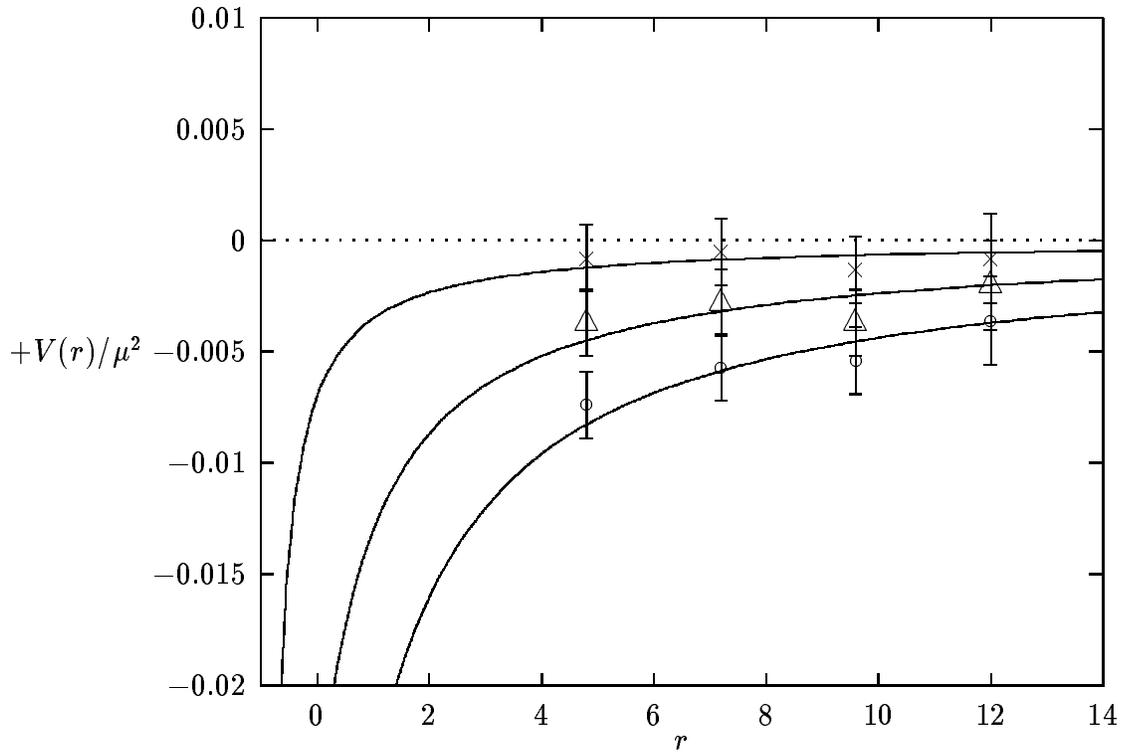

Fig. 7. Potential $V(r)/\mu^2$ on a linear scale, again for $\lambda = 1$ and $a = 0$, and $k$=0.03 ($\times$), 0.04 ($\triangle$), and 0.05 ($\circ$). ($k_c \approx 0.060$). The lines represent best fits to the data of the form $-c/r$ for $r > l_0 = 2.36$.

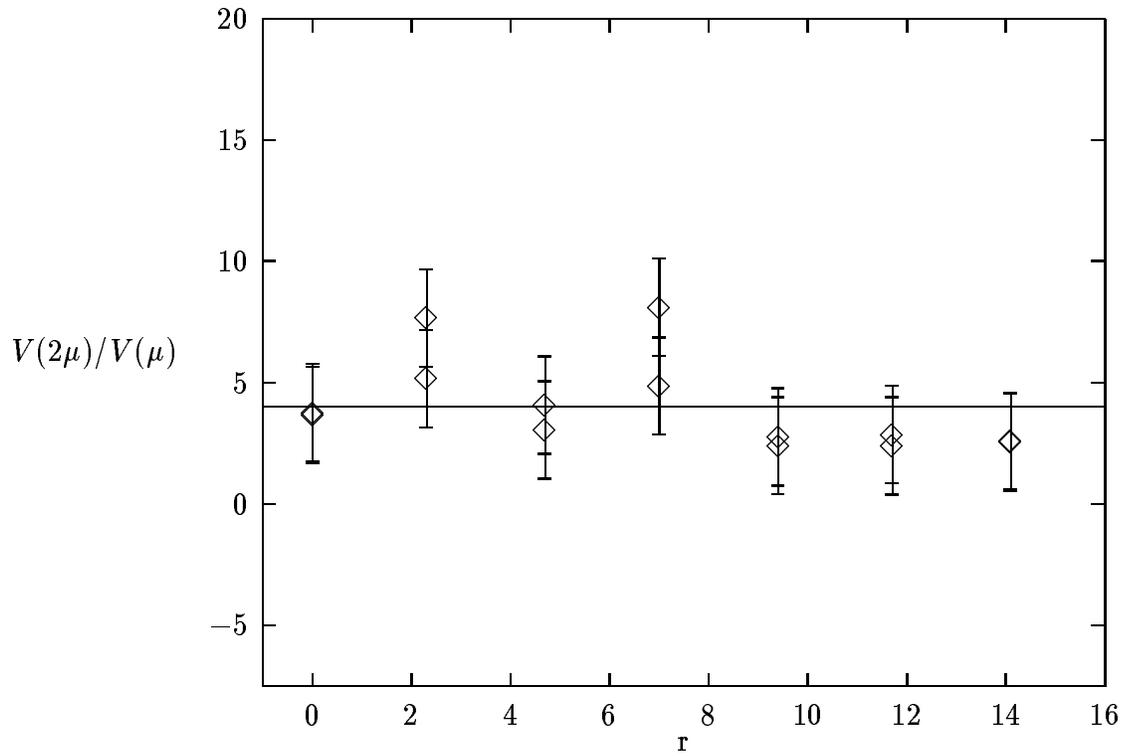

Fig. 8 . Ratios of potentials $V(2\mu)/V(\mu)$ for three different choices for the heavy masses ($\mu$=0.125,0.25,0.5) at $k = 0.04$. For a mass-squared dependence one expect the ratio to approach 4, independent of distance.

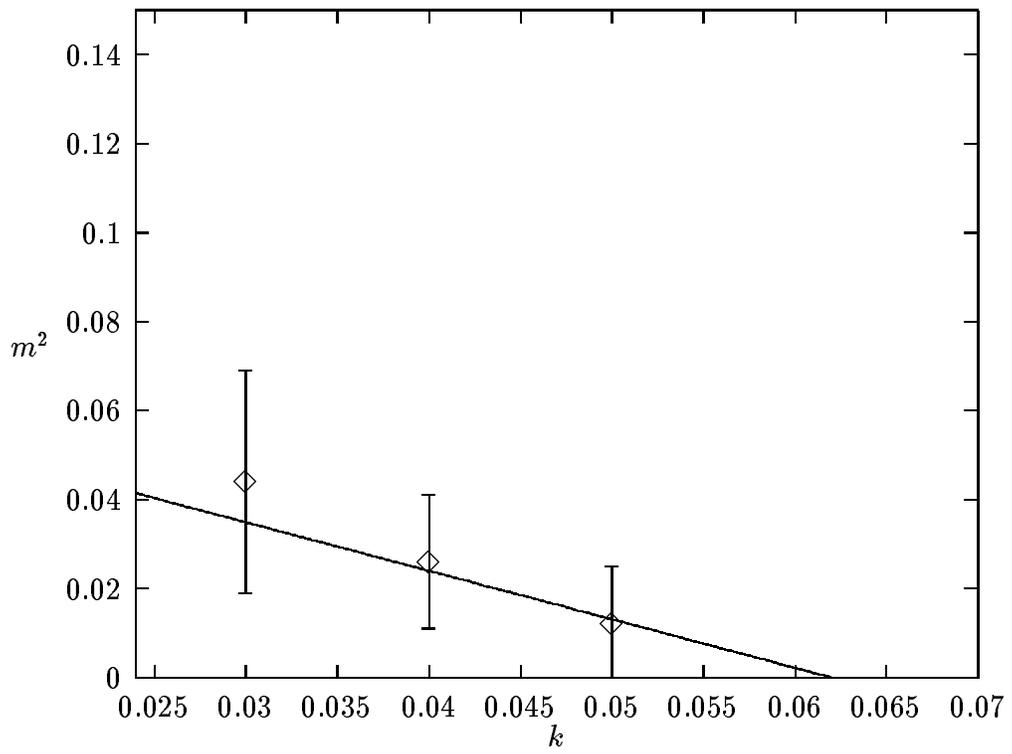

Fig. 9. Graviton mass parameter $m$ squared versus bare coupling $k = 1/(8\pi G)$.

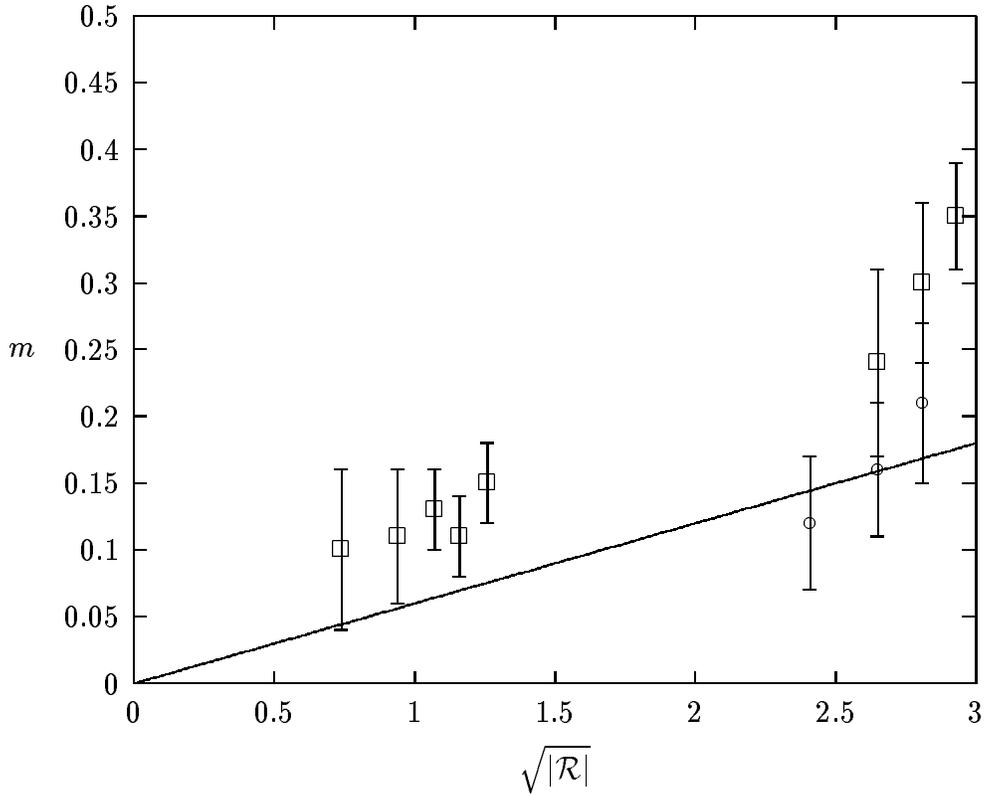

Fig. 10. Graviton mass parameter $m$ versus the average curvature $\mathcal{R}$ ($\circ$). For comparison we show the same mass parameter extracted from the invariant curvature-curvature correlations at fixed geodesic distance (from Ref. [47]) ($\square$) for $a = 0$ (points at large $\mathcal{R}$) and for $a = 0.005$ (points at small $\mathcal{R}$).

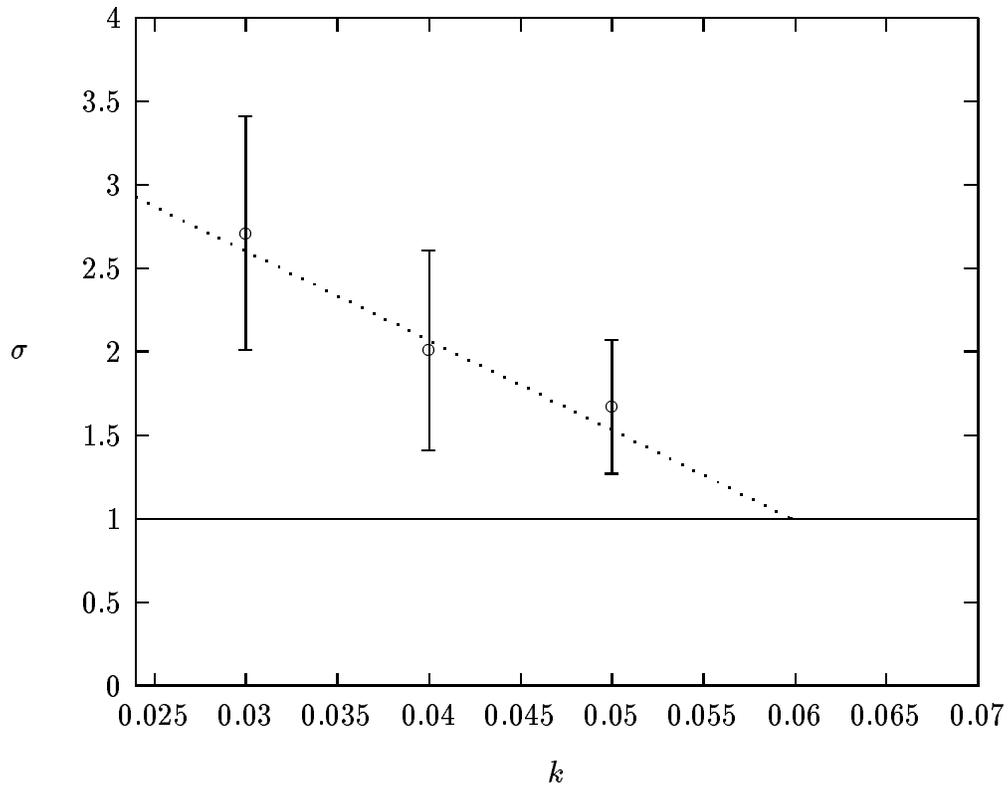

Fig. 11. Power $\sigma$ characterizing the decay of the potential versus bare coupling $k = 1/(8\pi G)$. The dotted line represents a linear fit, while the horizontal line corresponds to a $1/r$ dependence ($\sigma = 1$).

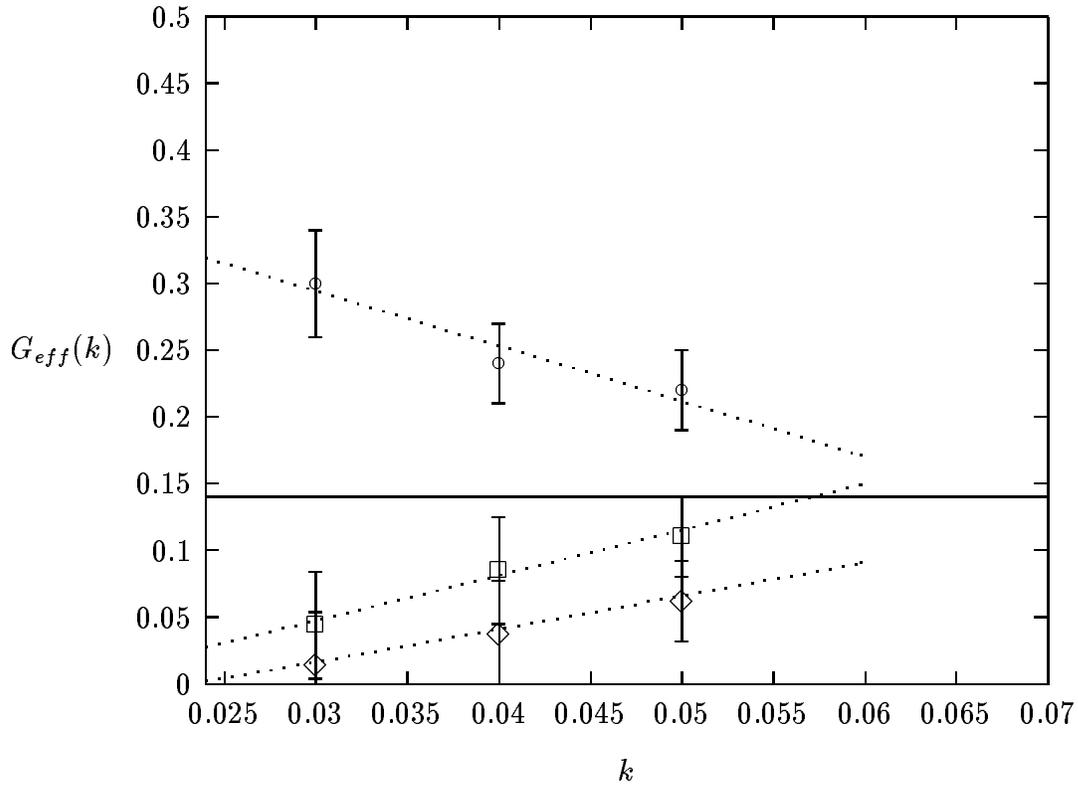

Fig. 12. Three methods of estimating the effective short distance Newton constant at the critical point. The effective Newton constant versus the bare coupling $k$ is computed using three different methods for extracting it; by fitting the potential to a form $c/r^\sigma$ ($\circ$), $c/r$ ($\square$), and $c\exp(-mr)/r$ ($\diamond$). The value estimated in the vicinity of the critical point at $k = k_c$ is represented by the horizontal line.